# PROCEEDINGS

## AEW11 – 2019

**11th- Asia-Europe Workshop on Concepts in Information theory**

**A Tribute to Hiro Morita (UEC, Chofu, Japan)**

**Royal Maas Yacht Club, Rotterdam, the Netherlands**
**July  3 - 5,  2019**

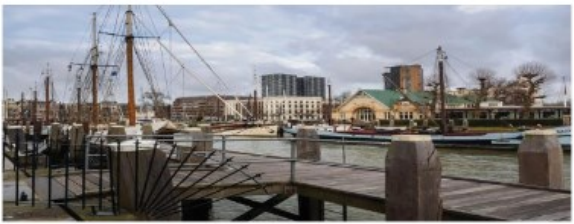 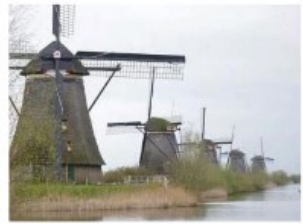 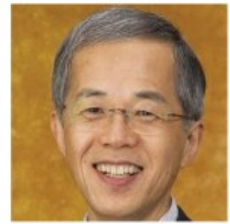

# PROCEEDINGS

AEW11-2019

11th Asia-Europe Workshop on Concepts in Information theory

A Tribute to Hiro Morita (UEC, Chofu, Japan)

Royal Maas Yacht Club, Rotterdam, the Netherlands

July 3 – 5, 2019

**Chairs**
　　　Kui Cai (SUTD, Singapore)
　　　Ludo Tolhuizen (Philips, Eindhoven)
　　　Hirosuke Yamamoto (Universityof Tokyo)

**Local Arrangements**
　　　Kees Immink (Turing Machines)
　　　JosWeber (TUD, Delft)

**Technical Program**
　　　Han Vinck (DUE, Duisburg)
　　　Tadashi Wadayama (NIT, Nagoya)

**International Advisory Committee**
　　　Jasper Goseling (UT, Twente)
　　　Hiroshi Kamabe (Gifu University)
　　　Vitaly Skachek (University of Tartu)

**Financial support**
　　　WIC, Werkgemeenschap voor Informatietheorie in de Benelux
　　　IEEE Benelux Chapter on Information Theory
　　　TURING MACHINES Inc

**Technical support**
　　　Shannon Foundation
　　　Stichting Leibniz

## Preface:

In 1989 we organized the first Benelux-Japan workshop on Information and Communication theory in Eindhoven, the Netherlands. This year, 2019 we celebrate 30 years of our friendship between Asian and European scientists at the AEW11 in Rotterdam, the Netherlands. Many of the 1989 participants are also present at the 2019 event. This year we have many participants from different parts of Asia and Europe. It shows the importance of this "small" event.

It is a good tradition to pay a tribute to a special lecturer in our community. This year we selected Hiroyoshi Morita. Hiro is a well known information theorist with many original contributions. We also appreciate very much his contributions to the Information theory community in general.

We expect all contributors to this workshop to pay special attention to the concept of their work. In this way, the workshop is also of interest to our young newcomers

The organizers prepared an ideal environment in the world-harbor city of Rotterdam, also known for its modern design architectures, for the exchange of ideas and intensive discussions.

A warm welcome from the organizers

K.A. Schouhamer Immink and A.J. Han Vinck

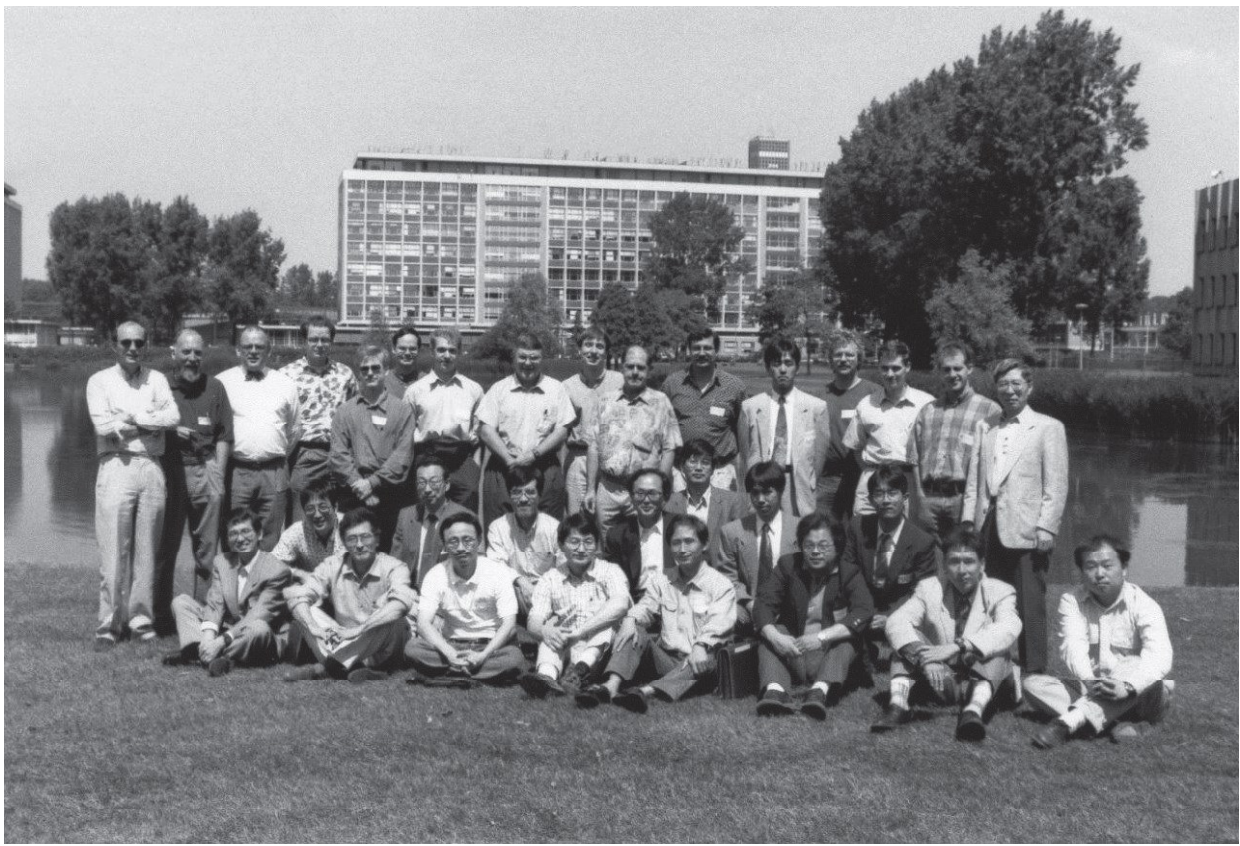

1st Japan-Benelux Symposium, 1989, Eindhoven, the Netherlands

# CONTENT:

# Data-Driven Tuning of Proximal Gradient Algorithms for Signal Recovery Problems

Tadashi Wadayama

Nagoya Institute of Technology, Japan

Email: wadayama@nitech.ac.jp

## I. Introduction

*Proximal gradient algorithms* [5] are iterative minimization algorithms for minimization of convex problems. In the field of sparse signal recovery, iterative soft thresholding algorithm (ISTA) is a well known proximal algorithm for solving Lasso problems. Recently, the author and his group have studied *data-driven tuning* for proximal gradient (and projected gradient) algorithms [2], [3], [4], [6], which are competitive (or even superior under certain conditions) to known signal recovery algorithms. The goal of this presentation is to survey the recent progress of our studies and present the underlying concept in these works.

The idea of data-driven tuning is concisely illustrated in Fig. 1. We can unfold the signal-flow of an iterative algorithm and obtain a multi-layer signal-flow graph which is similar to a feedforward deep neural network. It is expected that the behavior of each process is controlled by the trainable parameters (black circles in the figure). If each process of the signal-flow graph is differentiable, these trainable parameters can be adjusted by standard deep learning techniques. Namely, in order to optimize the internal parameters embedded in the iterative algorithm, we can use a "standard deep learning tool kit" such as back propagation and a stochastic gradient descent method. In a training process, the trainable parameters are tuned to minimize the loss function (e.g., squared error function) for a given training data set. This framework was originally proposed by Gregor and LeCun [1].

## II. Example

We present the idea of data-driven tuning with a simple example. Assume that the observation word $\boldsymbol{y} \in \mathbb{R}^n$ is given by $\boldsymbol{y} = \boldsymbol{A}\boldsymbol{x} + \boldsymbol{w}$ where $\boldsymbol{A} \in \mathbb{R}^{n \times n}$. A vector $\boldsymbol{x} \in \{+1, -1\}^n$ can be regarded as a transmitted vector to an AWGN channel and $\boldsymbol{w}$ is an additive Gaussian noise vector. The goal of the receiver is to estimate $\boldsymbol{x}$ from $\boldsymbol{y}$ as correct as possible. The problem is, in general, a computationally hard problem.

An approach to solve the estimation problem is to use a proximal gradient (PG) algorithm; we can use the following recursive formula to obtain the estimate:

$$\boldsymbol{r}_t = \boldsymbol{s}_t + \gamma_t \boldsymbol{A}^T(\boldsymbol{y} - \boldsymbol{A}\boldsymbol{s}_t), \quad (1)$$
$$\boldsymbol{s}_{t+1} = \tanh\left(\xi \boldsymbol{r}_t\right), \quad (2)$$

with an initial condition $\boldsymbol{s}_1 = \boldsymbol{0}$. There are two processes in the PG algorithm: In the gradient descent step (1), the vector $\boldsymbol{r}_t$ is updated along with the gradient descent direction of the objective function, i.e., $-\nabla \frac{1}{2}\|\boldsymbol{A}\boldsymbol{x} - \boldsymbol{y}\|_2^2 = \boldsymbol{A}^T(\boldsymbol{y} - \boldsymbol{A}\boldsymbol{x})$. The parameter $\gamma_t$ corresponds to the step-size parameter which controls the convergence behavior such as convergence to a fixed point and the convergence speed. In the proximal operation step (2), we apply a soft-projection function $\tanh(\cdot)$ to $\boldsymbol{r}_t$ in order to obtain the estimate $\boldsymbol{s}_{t+1}$ of the $t$th iteration step.

For a given iterative algorithm, there are a plenty of freedom to embed trainable parameters. A natural choice for (1)(2) is to choose $\gamma_t$ and $\xi$ as trainable parameters. In the presentation, contribution of data-driven tuning will be discussed through this example.

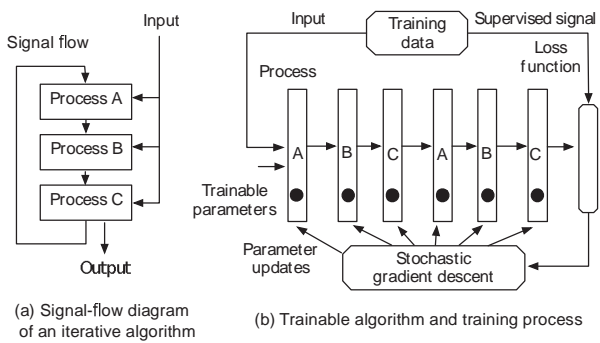

Fig. 1. Unfolded signal-flow graph with embedded trainable parameters.

# Weakly Constrained Codes and their Applications in Data Storage Systems

Van Khu Vu *

Nanyang Technological University, Singapore

Email: vankhu001@ntu.edu.sg

In coding theory and information theory, constrained codes have been studied actively for a long time with numerous applications in data storage systems, including compact disk [1], phase-change memory [2], flash memory [3], racetrack memory [4], and DNA-based storage [5].

In this work, we are interested in applications of constrained codes in flash memories, a popular non-volatile information storage technology. There are two main challenges to reliable data recording and retrieval in flash memories: namely inter-cell interference and charge leakage. Codes with local constraints, where some specific patterns are forbidden, are proposed to mitigate the inter-cell interference [3]. Moreover, codes with global constraints, in particular balanced codes or constant composition codes, can mitigate the charge leakage [6]. Our first goal is to study codes with global and local constraints to mitigate both inter-cell interference and charge leakage.

Constrained code is well-studied with many techniques to find the capacity and to encode/decode efficiently [7], [8]. Codes with both global and local constraints, for example balanced codes with run-length constraints [9] or balanced ICI-free codes, have been studied recently. We first investigate q-ary constant composition codes avoiding a set of specific patterns. Using enumerative techniques, we find the exact maximal size of these codes and thus compute the capacity of this constrained channel.

However, avoiding all the problematic patterns reduces the storage capacity. Hence, in the spirit of weakly constrained codes, proposed by Immink [10], we present the class of q-ary constant-composition weakly-constrained codes whose codewords have constant composition and allow some specific patterns to appear a certain number of times. As before, we derive results in terms of maximal size, asymptotic rates, and optimal composition ratio. We also present efficient encoding/decoding algorithms of these codes.

# On the Maximum Sum-Rates of Input-Constrained WOM Codes

Akiko Manada
*Dept. of Information Science*
*Shonan Inst. of Tech.*
1-1-25, Nishikaigan, Tsujido, Fujisawa
Kanagawa, 251-8511, JAPAN
amanada@info.shonan-it.ac.jp

Takahiro Ota
*Dept. of Computer & Systems Eng.*
*Nagano Prefectural Inst. of Tech.*
813-8, Shimonogo, Ueda
Nagano, 386-1211, JAPAN
ota@pit-nagano.ac.jp

Hiroyoshi Morita
*Grad. School of Informaics & Eng.*
*The Univ. of Electro-Communications*
1-5-1, Chofugaoka, Chofu
Tokyo, 182-8585, JAPAN
morita@uec.ac.jp

## I. INTRODUCTION

A Write Once Memory (WOM) is a storage medium consisting of level cells for which only transitions to higher levels are allowed when rewriting messages. A $t$-write WOM code is a coding scheme that can rewrite messages $T$ times only using the allowed transitions. For example, the first binary 2-write WOM code for 4 messages (say 0,1,2,3) using 3 cells, introduced by Rivest and Shamir [1], write messages as shown in Table I. Since then, many coding schemes and applications towards practical uses (e.g. flash memories) have been studied.

### TABLE I
2-WRITE WOM CODE FOR 4 MESSAGES USING 3 CELLS

|           | 0   | 1   | 2   | 3   |
|-----------|-----|-----|-----|-----|
| 1st write | 000 | 100 | 010 | 001 |
| 2nd write | 111 | 011 | 101 | 110 |

One of the main concerns on $t$-write WOM codes is the sum-rate which gives us the amount of information written in the $t$ writes. More precisely, a $t$-write WOM code of $n$ cells is determined as

$$[t, n : M_1, M_2, \ldots, M_t],$$

where $M_i$ ($1 \leq i \leq t$) represents the size of messages that can be stored at the $i$-th write, and the *sum-rate* $R_{sum}$ of the $t$-write WOM code is given by

$$R_{sum} = \frac{\sum_{i=1}^{t} \log_2 M_i}{n}.$$

Clearly, WOM codes with high sum-rates would be more desirable, and Fu and Vinck [2] precisely derive the maximum sum-rate of $q$-ary $t$-write WOM codes as follows.

**Corollary I.1** (Corollary 3.2 in [2])**.** *The maximum-sum rate* $R_{max}^{(q)}$ *for the* $q$-*ary* $t$-*write WOM codes is*

$$R_{max}^{(q)} = \sum_{i=1}^{q-1} \left[ \log_2(t+i) - \log_2 i \right]$$

## II. OUR CONTRIBUTIONS

It is sometimes more natural to consider additional constraints in practical sense (*e.g.* [3]–[5]). In this presentation, based on the work on Qin *et. al.* [4] we consider sum-rates of WOM codes when constraints on data sequences should be considered. More precisely, we first deepen our understanding of the technique in [4] to compute the maximum sum-rates of certain input-constrained WOM codes. To do so, we clarify the relationship between the maximum sum-rate of an input-constrained WOM code and the capacity of a two dimensional constrained system satisfying the input constraint. Indeed, it is important to recall that the capacity of a constrained system (see, for example, [4], [6]) is explicitly computed using the adjacency matrix of a graph representing the system.

Amongst various types of constraints, we focus, in particular, on the constraint to avoid on Inter Cell Interference (ICI), a typical error that occur when storing data in flash memories. It is known that ICI tends to be caused when there are big gaps between the symbol in the cell and the symbols in the neighbouring cells. For simplicity, we consider the binary case and set the input constraint as 101; that is, we assume that 101 cannot be appeared within data sequences. In this case, we present a matrix form to derive the maximum sum-rate.

## ACKNOWLEDGEMENT

This work was supported in part by JSPS KAKENHI, Grant Numbers 15K15936, 17K00147.

# Quantization for Emerging Non-Volatile Memories

Kui Cai

Singapore University of Technology and Design (SUTD)

8 Somapah Road, Singapore 487372

Email: cai_kui@sutd.edu.sg

## ABSTRACT

Channel quantization plays a critical role for high-speed emerging non-volatile memories (NVMs) such as the spin-torque transfer magnetic random access memory (STT-MRAM), where high-precision analog-to-digital converters (ADCs) are not applicable. We first investigate the design of the single-bit quantizer which is highly suitable for practical applications. In particular, we propose a quantized channel model for STT-MRAM, based on which we derive various information theoretic bounds for the quantized channel, including the channel capacity, cutoff rate, and the Polyanskiy-Poor-Verdú (PPV) finite-length performance bound. By using these channel measurements as criteria, we design and optimize the single-bit quantizer numerically for the STT-MRAM channel.

Next, in order to mitigate the unknown offset of the STT-MRAM channel caused by the variation of working temperature, we propose novel machine learning based dynamic detection schemes. We first present novel neural network (NN) based detectors that can effectively tackle the unknown offset of the channel. However, compared with the conventional memory sensing scheme, the NN detectors will incur a significant delay of the read latency and more power consumption. Therefore, we further propose a novel dynamic quantization scheme, where the corresponding quantization threshold can be derived based on the outputs of the NN detectors. In this way, the NN-based detection only needs to be invoked when the error correction code (ECC) decoder fails, or periodically when the system is in the idle state. Thereafter, the conventional memory sensing scheme will still be adopted by using the adjusted quantization threshold derived base on the outputs of the NN detector, until a further adjustment is needed. Simulation results demonstrate the effectiveness of our proposed channel quantization schemes for improving the data storage reliability of STT-MRAM.

# A Note on the Bayes Optimal Estimator of the Expected Intervention Effect

Shunsuke Horii

Waseda University

1-6-1, Nishiwaseda, Shinjuku-ku,

Tokyo 169-8050, Japan

Email: s.horii@aoni.waseda.jp

*Abstract*—In this work, we deal with the problem of estimating the expected intervention effect in the statistical causal analysis using the structural equation model and the causal diagram. The intervention effect is defined as an causal effect on the response variable $Y$ when the treatment variable $X$ is fixed to a certain value by an external agent. In general, there are three steps to estimate the causal effect: 1. Estimate the causal diagram from the data, 2. Estimate the conditional probability distributions in the causal diagram from the data, 3. Calculate the causal effect. However, if the problem of estimating the causal effect is formulated in the statistical decision theory framework, estimation with this procedure is not necessarily optimal. In the previous work, we obtained the Bayes optimal estimator of the causal effect when the loss function is the Kullback-Leibler divergence between the true causal effect and its estimator. In this work, we present the Bayes optimal estimator of the expected causal effect when the loss function is defined based on the expected causal effect and the estimator.

## I. Introduction

Causal analysis based on structural equation model is widely used in sociology, economics, biology, etc. Pearl defined a notion of causal effect named intervention effect [1]. Fixing a variable $X$ at a certain value $x$ by an external operation is called intervention, and the intervention effect is mathematically defined as a causal effect on the response variable $Y$:

$$p(y|\mathrm{do}(X=x), m, \boldsymbol{\theta}_m) = \int \cdots \int \frac{p(x, y, z_1, \ldots, z_p | m, \boldsymbol{\theta}_m)}{p(x | \mathrm{pa}(x), m, \boldsymbol{\theta}_m)} dz_1 \ldots dz_p. \quad (1)$$

In general, the calculation of the intervention effect based on the causal diagram consists of the following steps.

1) Estimate a causal diagram from the data
2) Estimate the conditional probability distributions among variables from the data
3) Calculate the intervention effect

However, if we formulate the problem of estimating the intervention effect based on the statistical decision theory, estimating it by this procedure is not necessarily optimal. In the previous study [2], we showed that the Bayes optimal estimator of the intervention effect under the Kullback-Leibler divergence loss is given by

$$AP^*(D^n) = p(y|\mathrm{do}(X=x), D^n), \quad (2)$$

where

$$p(y|\mathrm{do}(X=x), D^n) = \sum_{m \in \mathcal{M}} p(m|D^n) p(y|\mathrm{do}(X=x), m, D^n), \quad (3)$$

$$p(y|\mathrm{do}(X=x), m, D^n) = \int p(y|\mathrm{do}(X=x), m, \boldsymbol{\theta}_m) p(\boldsymbol{\theta}_m | m, D^n) d\boldsymbol{\theta}_m. \quad (4)$$

## II. Bayes Optimal estimator of the expected intervention effect

Sometimes, we are interested in the average of causal effects. For example, the average treatment effect (ATE) is the difference in average outcomes between units assigned to the treatment and units assigned to the control [3]. We define the expected intervention effect as follows.

$$\bar{y}_x = \int y \cdot (y|\mathrm{do}(X=x), m, \boldsymbol{\theta}_m) dy \quad (5)$$

Decision function $AP : D^n \mapsto \mathbb{R}$ outputs an estimate of the expected intervention effect. In this study, the quadratic loss is used as a loss function.

$$Loss(m, \boldsymbol{\theta}_m, AP(D^n)) = (\bar{y}_x - AP(D^n))^2. \quad (6)$$

Then, the Bayes optimal decision function is given by

$$AP^*(D^n) = \int y \cdot p(y|\mathrm{do}(X=x), D^n) dy. \quad (7)$$

## Acknowledgment

This research is partially supported by No. 19K12128 of Grant-in-Aid for Scientific Research Category (C) and No. 18H03642 of Grant-in-Aid for Scientific Research Category (A), Japan Society for the Promotion of Science.

# A Note on a Private PEZ Protocol

Yoshiki Abe, Mitsugu Iwamoto, and Kazuo Ohta

The University of Electro-Communications, Japan.
E-mail: {yoshiki,mitsugu,kazuo.ohta}@uec.ac.jp

*Abstract.* A private PEZ protocol is a variant of secure multi-party computation (MPC) by using a (long) PEZ dispenser. The original paper by Balogh et al. [1] presented a private PEZ protocol for computing arbitrary function $f_n$ with $n$-inputs. This result is interesting not only because MPC can be implemented by physical tools but also because the protocol requires no randomness for executing MPC. Unfortunately, however, no follow-up work has been presented since then, as far as the authors know.

In this paper, we show that it is possible to shorten the initial string (sequence of candies to be enclosed in a PEZ dispenser) drastically if the function $f_n$ is *symmetric*. Our main idea is to utilize the recursive structure of the symmetric function $f_n$ and construct the initial string recursively. As a result, it turns out that the length of initial string is reduced from $\mathcal{O}(2^n!)$ for general functions to $\mathcal{O}(n \cdot n!)$ for symmetric functions.

*Acknowledgment.* The authors are grateful to Mr. Shota Yamamoto for insightful discussion. This work was supported by JSPS KAKENHI Grant Numbers JP17H01752, JP18K19780, JP18K11293, JP18H05289, and JP18H03238.

# Neural Network-Based Dynamic Quantizer Design for LDPC-Coded STT-MRAM Channels

Xingwei Zhong, Kui Cai, and Zhen Mei

Singapore University of Technology and Design (SUTD), Singapore, 487372

## I. INTRODUCTION

Spin-torque transfer magnetic random access memory (STT-MRAM) is a promising emerging non-volatile memory (NVM) technology [1]. For STT-MRAM, a critical and difficult issue to be tackled is the memory physics induced unknown offset of the channel [1]. That is, with the increase of temperature, the low resistance of the STT-MRAM cell hardly changes, while the high resistance decreases, leading to more overlapping of the memory resistance distributions. The corresponding deviations from the nominal values of memory readback signals, called offsets, are unknown to the channel detector, and hence will severely degrade its error performance and lead to more decoding errors of the error correction code (ECC) subsequently. In [1], we present neural network (NN) based detectors that can effectively tackle the unknown offset of the channel. However, carrying out NN detection for every data block will incur a significant delay of the read latency and more power consumption. Although a dynamic threshold detector (DTD) based on the NN detection was further proposed in [1], it is essentially a one-bit quantizer and hence can only support the hard-decision decoding (HDD) of ECCs. For the soft-decision decoding (SDD) of ECCs such as the low-density parity-check (LDPC) codes for NVMs, more quantization bits are needed. In this work, we propose a novel NN-based dynamic design of the channel quantizer, which can generate soft reliability information of the channel coded bits and hence enable SDD of ECCs, without the prior knowledge of the channel.

## II. NEURAL NETWORK-BASED DYNAMIC QUANTIZER DESIGN

In order to minimize the number of quantization bits to reduce the area and latency of the memory sensing circuit, channel quantization should be concentrated on the less reliable region(s) that the channel readback signal falls in [2]. For the STT-MRAM channel with unknown offset, we propose to make use of the outputs of the NN detector to effectively and dynamically locate the least reliable region of the channel readback signal. In particular, based on the outputs of the NN detector, we can generate the hard estimate $\hat{x}$ of the channel input bits. Thereafter, for given channel readback signals $y$, we can search for a detection threshold $\tilde{R}_{th}$ that minimizes the Hamming distance between $\hat{x}$ and the channel detection bits $\bar{x}$ obtained by comparing $y$ with the corresponding $\tilde{R}_{th}$. The obtained $\tilde{R}_{th}$ actually indicates the least reliable region of the channel readback signal, and hence more quantization levels should be assigned around it. Therefore, in the proposed dynamic design of the quantizer, we consider a $q$-bit quantizer with $L = 2^q$ quantization levels. Let $t_0, t_1, \ldots, t_L$ with $t_0 = -\infty$ and $t_L = +\infty$ be the boundaries of the quantiza-tion intervals. We let $t_{L/2} = \tilde{R}_{th}$, and uniformly quantize the region $[t_1, \ldots, t_{L-1}]$ such that the capacity of the quantized channel is maximized [2]. The integer index of the quantization interval that the readback signal belongs provides the reliability information of the corresponding channel input bit, and hence can be sent to the soft-decision decoders, such as the reliability-based min-sum (RB-MS) decoder of LDPC codes. Simulation results illustrated by Fig. 1 demonstrate that the proposed dynamic quantizer design (with $q = 3$) based on the RNN detection can approach the error rate performance of the quantizer designed with the full knowledge of the channel.

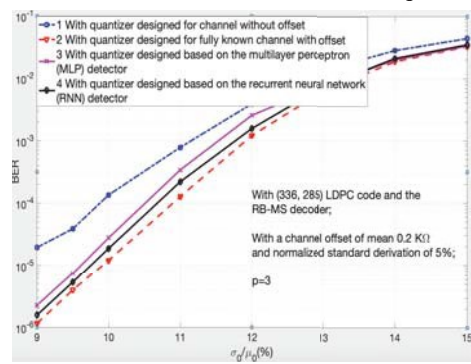

Fig. 1. Decoding BERs with different quantizers.

# Comparisons of Detection Schemes for Channels with Unknown Varying Offset

Renfei Bu                    Jos Weber

Delft University of Technology

Applied Mathematics Dept., Optimization Group

Van Mourik Broekmanweg 6, 2628 XE Delft, The Netherlands

`R.Bu@tudelft.nl`          `J.H.Weber@tudelft.nl`

Pearson distance-based detection [1] has been proposed to counter the effects of gain and/or offset mismatch in noisy channels, where the gain and the offset may change from word to word, but are constant for all transmitted symbols within a codeword. Here, we consider the situation where the offset varies linearly within a codeword. A detection technique for such channels is investigated in [2], where Pearson distance-based detection is used in conjunction with mass-centered codes. The use of Pearson distance-based detection in cooperation with a difference operator is also shown to be immune to gain and varying offset mismatch [3]. In [3], pair-constrained codes are proposed for unambiguous decoding, where in each codeword, certain adjacent symbol pairs appear at least once.

We investigate performances of these two detection schemes in terms of: (1) noise resistance ability; (2) cardinality and redundancy. The noise performance of the detection scheme with mass-centered codes is better than its counterpart that uses a difference operator. On the other hand, the redundancy of pair-constrained codes is much lower than that of mass-centered codes, which makes the detection scheme in [3] an attractive alternative for practical applications.

# Codes for Load Balancing in Multi-Server Systems

**Vitaly Skachek**
Institute of Computer Science
University of Tartu, Tartu 50409, Estonia
E-mail: `vitaly.skachek@ut.ee`

EXTENDED ABSTRACT

Consider a distributed database storing data on $n$ servers, and assume that the users query data at high rate. Typically, many users request a particular popular subset of data, called *hot data*. However, in that scenario, load on the servers is not evenly balanced, and the server that stores the hot data may become a bottleneck in the system.

In order to overcome this limitation, Ishai *et al.* [2] propose codes for load balancing, which are called *batch codes*. In that solution, the servers store coded data, and an user query is answered by reading data from a small number of servers. Yet, the same data can be read from a number of disjoint sets of servers.

**Definition 1** ([8])**.** *An $(n, k, t)$ batch code $\mathcal{C}$ over a finite alphabet $\Sigma$ is defined by an encoding mapping $\mathsf{C} : \Sigma^k \to \Sigma^n$, and a decoding mapping $\mathsf{D} : \Sigma^n \times [k]^t \to \Sigma^t$, such that*

1) *For any $\boldsymbol{x} \in \Sigma^k$ and $i_1, i_2, \cdots, i_t \in [k]$ ,*

$$\mathsf{D}\left(\boldsymbol{y} = \mathsf{C}(\boldsymbol{x}), i_1, i_2, \cdots, i_t\right) = (x_{i_1}, x_{i_2}, \cdots, x_{i_t}).$$

2) *The symbols in the query $(x_{i_1}, x_{i_2}, \cdots, x_{i_t})$ can be reconstructed from $t$ respective pairwise disjoint recovery sets of symbols of $\boldsymbol{y}$ (the symbol $x_{i_\ell}$ is reconstructed from the $\ell$-th recovery set for each $\ell$, $1 \leq \ell \leq t$).*

Take $\mathbb{F}$ to be a finite field $\mathbb{F}_q$, $q$ is a prime power. Let $\mathcal{C}$ be a linear $[n, k]$ error-correcting code over $\mathbb{F}$. Then, the encoding mapping $\mathsf{C}$ is given by:

$$\boldsymbol{y} = \boldsymbol{x} \cdot \boldsymbol{G} \in \mathbb{F}^n \, , \tag{1}$$

where $\boldsymbol{G}$ is a $k \times n$ generator matrix over $\mathbb{F}$, and $\boldsymbol{x} \in \mathbb{F}^k$ is an information vector. We use the term *linear batch code* (with parameters $n$, $k$ and $t$) over $\mathbb{F}$ for such a linear code, which supports any $t$ queries, and denote it as an $[n, k, t]_q$-*batch code* (see [3]). For a comprehensive survey about the properties of batch codes the reader can refer to [7]. A study of achievable service rates under the batch code model is presented in [1].

It is shown in [2] that there exist batch codes of rate $k/n \to 1$ for $k \to \infty$. Thus, it is interesting to obtain estimates on the optimal value of the redundancy $\rho(k) \triangleq n - k$. As it is shown in [4] and [8], for $t \in \{3, 4\}$ we have that $\rho(k) = \Theta(\sqrt{k})$, and for $t \geq 5$ we have $\rho(k) = O(\sqrt{k} \log k)$.

This work is supported in part by the Estonian Research Council grant PRG49, and by the ERDF through the Estonian Center of Excellence in ICT Research (EXCITE).

One limitation of the batch code model is that the queries are served in parallel, in a synchronous way. In reality, the queries may arrive in arbitrary times, and it could be desirable to start serving the existing queries before all $t$ queries have been received. In [6], a modified version of batch codes is given, which is referred to as *asynchronous batch codes*.

**Definition 2** ([6])**.** *An asynchronous linear $[n, k, t]_q$ batch code $\mathcal{C}$ over $\mathbb{F}$ is a linear batch code with the additional property that for any legal query $(x_{\ell_1}, x_{\ell_2}, \cdots, x_{\ell_t})$, all $\ell_i \in [k]$, it is always possible to replace $x_{\ell_j}$ (for $j \in [t]$) by some "new" $x_{\ell_{t+1}}$, $\ell_{t+1} \in [k]$, such that $x_{\ell_{t+1}}$ is retrieved from the servers not used for retrieval of $x_{\ell_1}, x_{\ell_2}, \cdots, x_{\ell_{j-1}}, x_{\ell_{j+1}}, \cdots, x_{\ell_t}$.*

It is observed in [6] that the codes constructed from bipartite graphs without short cycles [5] are asynchronous. By using the techniques from the area of extremal graph theory, it is shown in [6] that the optimal redundancy $\rho(k)$ of such graph-based asynchronous batch codes with $t = 3$ is $2\sqrt{k}$. Moreover, for a general fixed value of $t \geq 4$, $\rho(k) = O\left(k^{1/(2-\epsilon)}\right)$ for any small $\epsilon > 0$. For a general value of $t \geq 4$, $\lim_{k \to \infty} \rho(k)/\sqrt{k} = \infty$.

ACKNOWLEDGEMENTS

This work is based on collaborations with H. Lipmaa, S. Mikelsaar, Ü. Reimaa, A.-E. Riet, M. Simisker, E.K. Thomas, and H. Zhang.

# Overview: Antidictionary and its Applications

## Hiroyoshi Morita

University of Electro-Communications
Chofu, Tokyo 182–8585, JAPAN
Email: morita@uec.ac.jp

## Abstract

An antidictionary is the set of all words of minimal length that never appear in a given sequence of letters, or text. It was first introduced in information theory society by Chrochemore et al. in 2000 as a core mechanism for lossless data compression. After their landmark paper, its interests has been grown greatly in various research topics, such as frame-synchronization in digital communications, reconstruction of genomes in bioinformatics, and ECG anomaly detection in healthcare.

The task for building an antidictionary for a text $T$ is to find all particular words satisfying;

i)   they never occurs in $T$, and

ii)   their proper sub-words, or factors, do occur in $T$.

These words are called Minimal Forbidden Words, or MFW for short.

There have been many algorithms to find MFWs on $T$. As far as I know, in all of them, some tree data structures such as suffix trie, suffix tree, suffix automaton, and suffix array were utilized to get a clue to MFWs.

In this tutorial talk, I aim to explain how to find MFWs from a given text using suffix trie since it is the most comprehensible data structure to understand the characteristics of MFWs and also give an overview of other algorithms to find MFWs from a point view from suffix trie. Moreover, I will report some results on my recent works on ECG abnomaly detection using antidictionary.

This work is partially supported by JSPS KAKENHI Grant Number JP17K00400.

# Time Series Quantization and Compression subject to Distortion Measures

A.V. Poghosyan[1], A.N. Harutyunyan[1], N.A. Hovhannisyan[1], N.M. Grigoryan[1],
A.J. Han Vinck[2,3], and Y. Chen[2]
[1]VMware Eastern Europe
[2]Institute of Digital Signal Processing, University of Duisburg-Essen
[3]University of Johannesburg

### Abstract

Efficiently storing storms of data to preserve their utility is a challenging data science problem in many technology applications such as management of cloud computing infrastructures. The problem is information-theoretic in nature. We consider various compression approaches (which are based on Machine Learning) to time series (metric) data measured from IT or cloud computing resources (virtual machine, host, application server, etc.) while monitoring those distributed systems. The main idea is to apply time series quantization and then compress the data with elimination of sequential duplications of values.

## Introduction

Reliable management of modern cloud infrastructures and applications are increasingly relying on monitoring of massive volumes of time series data related to every aspect of the data center and storing them in specific data structures. This leads to a quite large storage consumption problem if considering also a vast number of self-generated time series needed for achieving the management goals. Commonly, it requires terabytes of storage space for archiving months/years of monitored data from IT resources/objects, each with hundreds/thousands of parameters (measured as individual time series metrics). We need a procedure to optimize data storage and information retrieval processes, especially for users/customers who need to store years of data using "realistic" storage space and disk I/O utilization requirements. Direct implementation of well-known compression methods will not solve the problem as some strict practical limitations exist:

a) Maintaining the common "time stamp - data value" structure of time series. Hence, we can't use, for example, vocabulary-based techniques for lossless compression or approximations, interpolations, and transformations for lossy compression;

b) Although, the storage space and I/O minimization have the highest priority, in general, a compromise between those minimizations and time of compression-decompression, CPU, Memory utilizations is a challenge. Hence, archiving approaches and approaches with heavy decompression are unacceptable.

We suggest a generic approach in line with the above-mentioned limitations – quantization of time series (lossy operation) and elimination of data value repetitions (lossless operation). This approach would be effective if the quantized time series is a low-variability data with big number of sequential data value repetitions. Quantization of time series is performed subject to distortion criteria imposed by users and preserving the main information content "enough" for effective management of the system. Distorted time series data can still be used for performing some analytical tasks such as outlier detection of new observations.

Below we discuss three different Machine Learning approaches (see [1,2]), targeting distinct use cases and complexity vs accuracy trade-offs. *Algorithm A* describes the process of quantization that maximizes the compression rate or minimizes the loss function. *Algorithm B* performs quantization with distortion depending on the data point importance – with lower distortion for important patterns and higher for the remaining. *Algorithm C* describes ideas for multi-variate time series with application of clustering methods. Related researches (for instance, [3]) in the domain of time series compression propose spline approximation-based algorithms. Our prior and related works include papers [4]-[8] linking also to relevant information-theoretic concepts.

## Algorithm A
## Optimal Quantization subject to Fidelity and Compression Rate

Under this scenario, the range of time series data is partitioned according to a set of quantiles. The quantized time series data is obtained from the original metric by substituting its values with the nearest quantiles. The quantized time series data is then compressed by removing sequential duplicate values. More specifically, quantization is performed by partitioning the metric range according to a selected number $n$ of quantiles denoted by $q_1, \dots, q_n$. The set of quantiles $\{q_i\}_{i=1}^{n}$ divides the data range of time series data into $n + 1$ groups of data points based on their values. Each group contains almost the same number of data points.

Let $x_k = x(t_k), k = 1, \dots, N$, be data points of a time series. Denote $X = \{x_k\}_{k=1}^{N}$. And let $x_k^q = x^q(t_k), k = 1, \dots, N$, be the nearest quantiles of the metric values corresponding to time stamps $t_k$. The sequence of quantized time series is represented by $X^q = \{x_k^q\}_{k=1}^{N}$.

Accuracy of this quantization can be measured, for instance, by $\ell_1$-error (loss function)

$$l_1 = \frac{1}{N} \sum_{k=1}^{N} |x_k - x_k^q|.$$

Compression of the quantized metric is then performed by elimination of sequential repetitions, or consecutive duplications, of quantized data points $x_k^q$ from the quantized data set $X^q$. Hence, we get a loosely compressed metric $x_k^c = x^c(t_k), k = 1, \dots, M$, where $M \le N$. Denote $X^c = \{x_k^c\}_{k=1}^{M}$.

The compression rate is given by

$$CR = 100 \frac{N-M}{N}.$$

The number of quantiles, $n$, is ideally selected to minimize the loss function and to maximize the compression rate.

However, simultaneous minimization of the loss function and maximization of the compression rate are contradicting operations.

Therefore, we consider two optimization set-ups. In the first set-up, the user requirement on the quality of compressed data is controlled by an upper bound ($\Delta$) on the loss function (LF), while the objective is to maximize the compression rate:

$$\begin{aligned} CR &\rightarrow max \\ LF &\le \Delta. \end{aligned} \qquad (1)$$

According to the second optimization setting, the user is interested in at least $r\%$ of storage reduction subject to minimum loss:

$$\begin{aligned} CR &\ge r \\ LF &\rightarrow min. \end{aligned} \qquad (2)$$

Optimization problems in (1) and (2) may result in different optimal number of quantiles $n$ to work with.

Figures 1 and 2 consider a specific infrastructure health metric and its quantization with four quantiles. Here, $n = 4$, $q_1 = 25$ (0.06-th quantile), $q_2 = 33$ (0.2-th quantile), $q_3 = 73$ (0.38-th quantile), $q_4 = 78$ (0.66-th quantile), $\ell_1/mean(data) = 0.026$ and $CR = 98.6\%$.

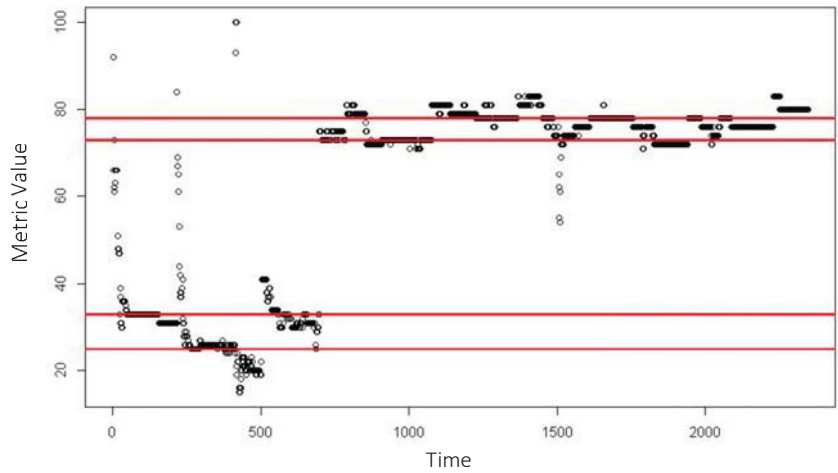

Fig. 1. Red solid lines show the quantiles which minimize the $\ell_1$ error.

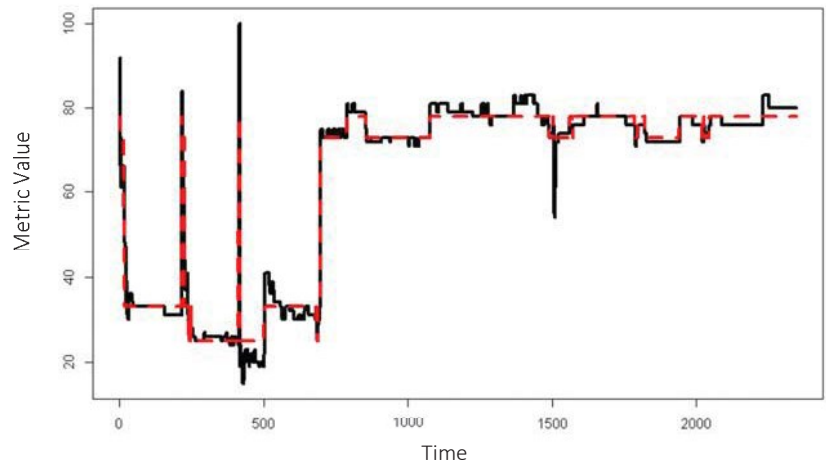

Fig. 2. Original (black solid) and quantized (red dashed) time series for $n = 4$, where the corresponding quantiles minimize the $\ell_1$ error.

**Algorithm B**
**Compression based on Data Importance Patterns**

Algorithm A quantizes the time series subject to a distortion measure. Assume that the data values in the metric are associated with different degrees of "importance" (can be treated based on a cost function). For example, data points from cluster "A" have "high importance", values from cluster "B" have "moderate importance", and values from cluster "C" have "low importance". Then, the compression is performed linked to the importance cost of the cluster. For example, we can apply lossless compression for the most important cluster "A" and lossy compression or lower accuracy compression for "B" and "C".

In general, importance of data points relates to behavioral patterns of time series data and can be different while working with performance, capacity, or configuration metrics. For example, in case of performance metrics, the importance of a data point relates to its "participation" in an anomaly process. For identifying a performance anomaly, dynamic or hard thresholding (see [9]-[11]) techniques can be applied. Data points that violate thresholds should be more important than data points within thresholds. Hence, while compressing historical data with known anomaly patterns, it is natural to preserve those anomalies (outliers) and largely compress/reduce within-thresholds data points by acceptable low accuracy.

Now, assume that $H$ and $L$ are upper and lower thresholds, respectively. Our compression approach is summarized in the following 5 steps:

**Step 1.** Determine the points satisfying the condition $x_k > H$ and store them without distortion with their corresponding time stamps.

**Step 2.** Determine the points satisfying the condition $x_k < L$ and store them without distortion with their corresponding time stamps;

**Step 3.** According to the required accuracy (say, user defined), determine the value of parameter $n$ indicating the number of slices the interval $[L, H]$ should be divided into by the following straight lines:

$$c_k = L + \frac{H-l.}{n}k, k = 0, \ldots, n.$$

**Step 4.** Determine the points in the intervals $I_k = [c_k, c_{k+1}), k = 0, \ldots, n-2, I_{n-1} = [c_{n-1}, c_n]$, and calculate the median, the average or any other reasonable statistical measure for each group of data points within $I_k$. Denote those measures by $m_0, \ldots, m_{n-1}$. If a data point belongs to the $I_j$ interval, then change its value to $m_j$. This results in a quantized time series with distorted in-bound and exact out-bound data points;

**Step 5.** Eliminate data value duplications with the corresponding time stamps. This results in a compressed time series. More duplications imply higher compression rates.

Let $v_k = x_{k+1} - x_k, k = 1, \ldots, N-1$, be the corresponding variability metric and V=$\{v_k\}$. We define variability measure (VarM) of a time series as a percentage of jumps in $X$

$$VarM = 100 \frac{Number\ of\ Non-Zero\ Components\ in\ V + 1}{Number\ of\ All\ Components\ in\ X}$$

$VarM = 0\%$ characterizes constant time series $X$ with $x_k \equiv c$. Conventionally, if $VarM \leq 50\%$, we deal with a low-variability metric, otherwise it is a high-variability metric.

$VarM$ can be determined both for original and compressed time series, $VarM(Original)$ and $VarM(Quantized)$, respectively. $VarM(Quantized)$ characterizes the compression rate (CR) of the above described algorithm defined as $CR = 100 - VarM(Quantized)$.

We see that the efficiency of our approach is strictly relates to the variability of the quantized time series. Of course, this approach can be applied immediately to a time series without quantization, but it will lead to poor results for high-variability metrics.

Figures 3 and 4 show the compression process of a high-variability time series ($VarM = 99.4\%$) while dividing within Dynamic Thresholds (DT – the typical time-varying ranges of time series, see [10]) area into $n = 2$ equal parts. In this case, the quantized time series is a low-variability data with $VarM = 33.2\%$. After elimination of duplications we get compression rate $CR = 66.8\%$ and $\ell_1 = 0.08$.

Now we describe the compression process for an entire database containing 86725 time series (metrics) from 309 different objects (VMs). Each of those time series have 1000-9300 data points. We found that 63.3% of those metrics are low-variability and 36.6% are high-variability (see Fig. 5).

Fig. 6 shows compression rates of high-variability data when $n = 10$. Compression rates of low-variability metrics range from 50% to 100% and are mostly concentrated around 100%. Fig. 7 demonstrates relative (divided over the averages) $\ell_1$-errors of quantization of high-variability time series. They are concentrated around 0.02. Similar errors of low-variability metrics are ranging from 0 to 0.04 but mostly are concentrated around 0.

Fig. 8 indicates average compression rates of different metric categories vs different number of within-DT segments. We see that compression rates of high-variability metrics are very sensitive to parameter $n$ and relatively acceptable values of $n$ are $n = 1,2,3$.

Fig. 9 graphs the average errors for different metric categories vs different number of within-DT segments. Overall, the trade-off between compression rate and error of quantization is achieved when $n = 1,2,3$.

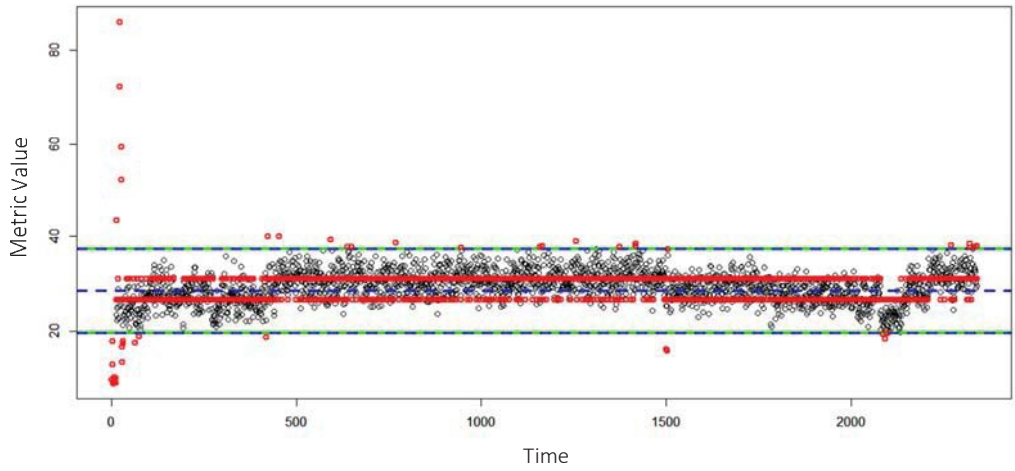

Fig. 3. Original high-variability time series with upper and lower dynamic thresholds (green solid lines). Red points show the values of the quantized time series.

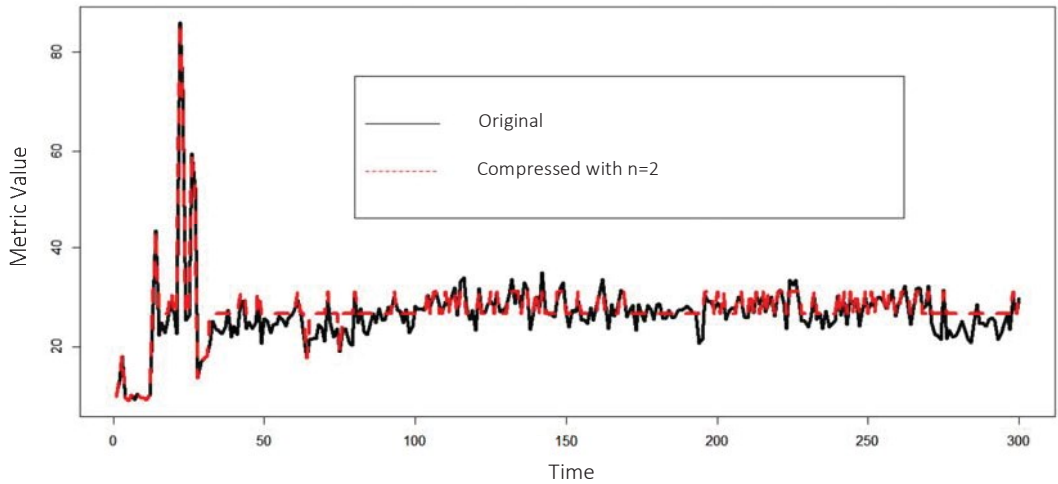

Fig. 4. Original high-variability time series (black solid) and quantized low-variability time series (red dashed).

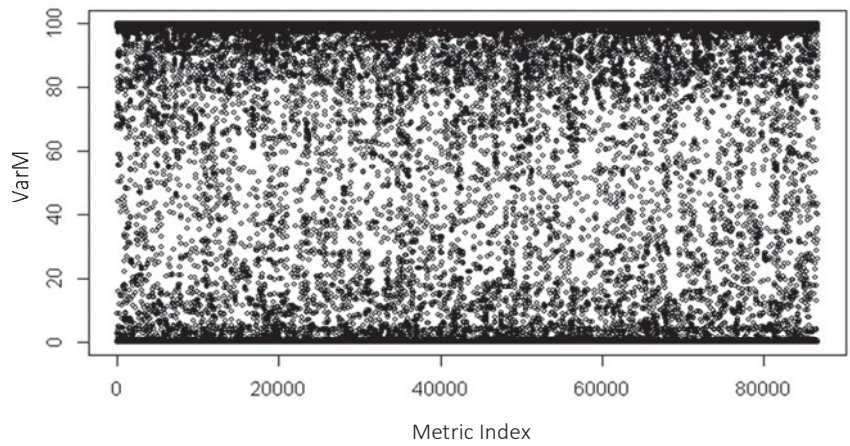

Fig. 5. Variability measures for different time series.

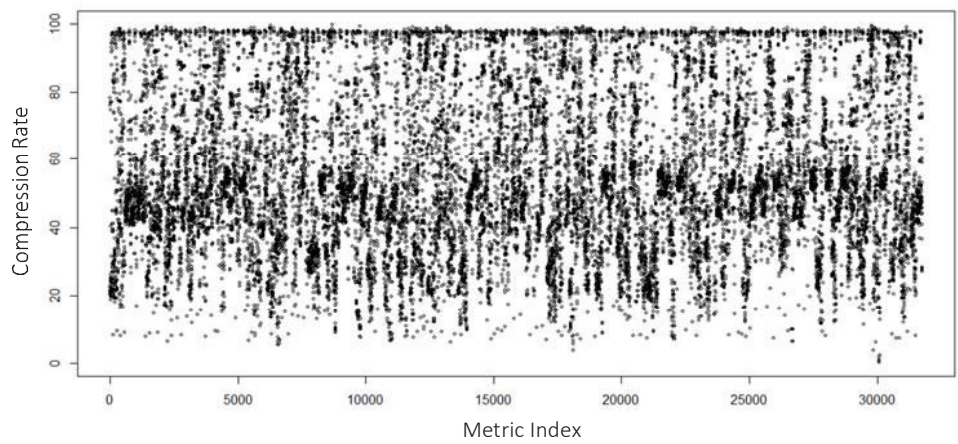

Fig. 6. Compression rates for high-variability metrics when $n = 10$.

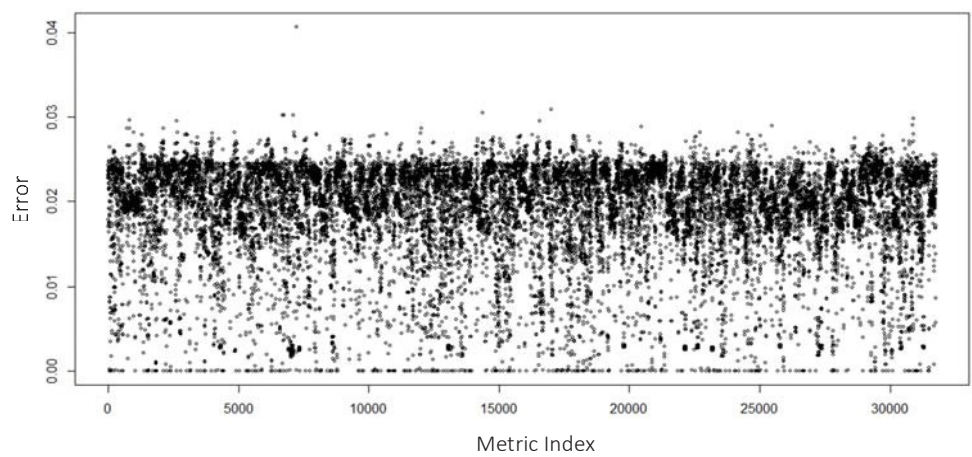

Fig. 7. Relative $\ell_1$-errors after quantization for high-variability metrics when $n = 10$.

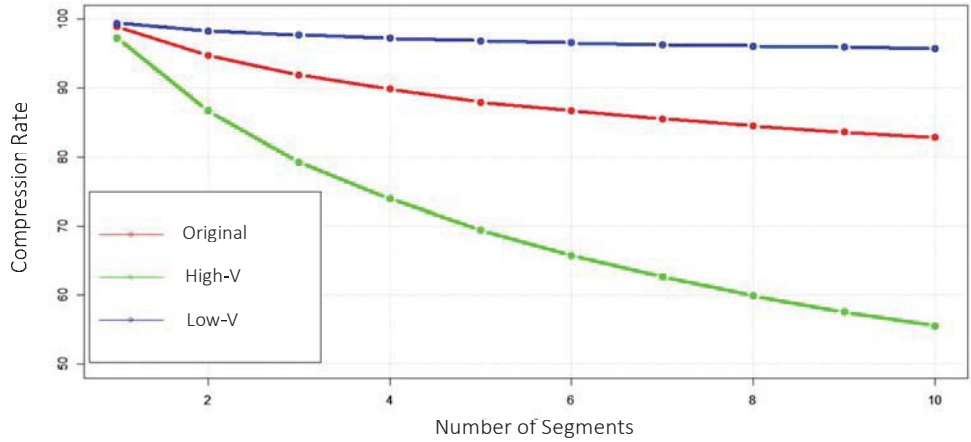

Fig. 8. Average compression rates for different metrics categories vs different number of within-DT segments.

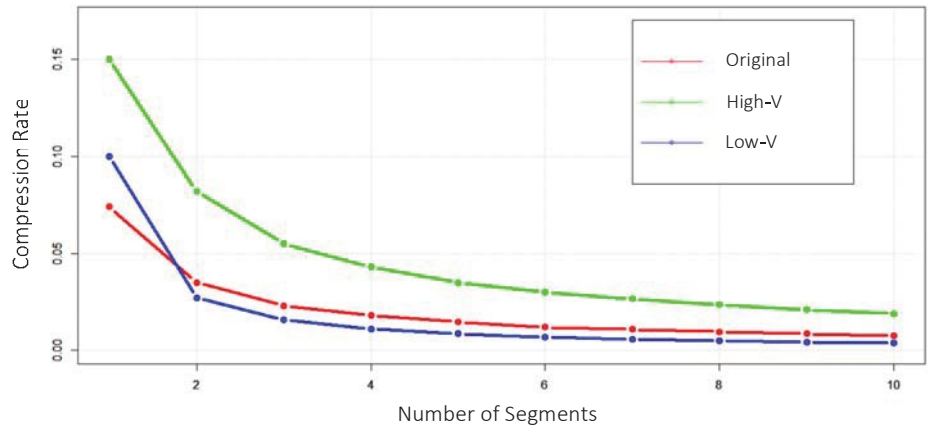

Fig. 9. Average errors for different metric categories vs different number of within DT segments.

### Algorithm C
### Multidimensional Data

Time series metrics can be aggregated into a multi-dimensional representation to analyze their trade-off behavior using machine learning/clustering. In other words, the values of $n$ different metrics at time stamp $t$: $m(t) \equiv (m_1(t), m_2(t), ..., m_n(t)) \in R^{n+1}$, make a point $m(t)$ in $(n+1)$-dimensional space, where one of the axes is the time.

In many cases, the time can be excluded from the multi-dimensional representation making a point $m_k$ in $n$-dimensional space $m = \{m_k\}_{k=1}^N$, $m_k = (m_{1,k}, m_{2,k}, ..., m_{n,k}) \in R^n$. We want to evaluate the historically typical trade-off between those metrics by analyzing the above-mentioned vectors. This leads to application of machine learning techniques of clustering ([12,13]) to identify typical spaces $m(t)$ or $m$ stays within.

How to store the high volumes of historical multidimensional data in a reduced way that it still contains underlying utility or "important patterns"? In particular, how to optimally compress the above-mentioned multi-variate time series data subject to a required fidelity measure $d$?

Such a goal statement implies an optimization problem of finding the best "Δ-coverage" of the data set according to the fidelity or distortion measure $d$. This coverage is defined by a subset of vectors

$$m_k = \{m_{1,k}, m_{2,k}, \ldots, m_{n,k}\}, k = 1, \ldots, K$$

such that for every other $m_k$ from the data set there is a point $m_{i_0}, 1 \leq i_0 \leq K$ satisfying the condition $d(m_k, m_{i_0}) \leq \Delta$.

The Δ-coverage with size $K$ is called optimal if there is no other coverage of lesser size. In an implementation, $d$ can be the Euclidean distance together with the $K$-means clustering approach:

1) to find the $K$ centroids of the data set;
2) verify if the distortion requirement $\Delta$ is met for all clustered points;
3) if it is not met, increase $K$ by 1 and repeat the procedure until it is satisfied.

As soon as the final centroids are found and the optimal coverage is achieved, our data reduction coding procedure collapses all within-cluster data points into corresponding centroids, so we store only those centroids and thus preserve the required fidelity in the data representation.

This reduced data set can be still used to perform streaming compression of newly arrived observations into one of the closest centroids ($k$-nearest neighbor logic) subject to distortion level. If that is not possible, then compose a new centroid and proceed further.

Above described procedure might lead to an unacceptable big value for $K$ in case of some outliers in multi-dimensional data. We consider an outlier-Δ-coverage assuming to preserve outliers exactly and apply optimal Δ-coverage for the remaining data points. This will help to achieve the same distortion with lesser number of clusters or better accuracy with the same coverage.

We selected two metrics from our private environment. One is "Badge-Health-Classic" taken as the first component of $m_k$ and the other is "CPU-0-Ready-Summation" taken as the second component (see Fig. 10). Then we performed $K$-means clustering with different $K \geq 1$ number of clusters. Assume that $c_j = (c_j^1, c_j^2)$ is the $j$-th cluster centroid. If a point $(x_k, y_k)$ belongs to the $j$-th cluster, we encode its value as $(c_j^1, c_j^2)$. Later, for each $K$, we calculate $\ell_{max}$-error (see Fig. 11). Formal definitions are as follows. Let $m_k = (x_k, y_k), k = 1, \ldots, N$, and distances are measured as follows:

$$d_1\left(m^{(1)}, m^{(2)}\right) = \max_k \left(\left(x_k^{(1)} - x_k^{(2)}\right)^2 + \left(y_k^{(1)} - y_k^{(2)}\right)^2\right)^{1/2}$$

or $d_2\left(m^{(1)}, m^{(2)}\right) = \underset{k}{\text{mean}} \left(\left(x_k^{(1)} - x_k^{(2)}\right)^2 + \left(y_k^{(1)} - y_k^{(2)}\right)^2\right)^{1/2}$, where $m_k^{(s)} = \left(x_k^{(s)}, y_k^{(s)}\right), s = 1,2$. It is natural to use relative measures, the results for different metrics can be compared easily. If $m_k^{(1)}$ is the original data and $m_k^{(2)}$ is its quantized version, then

$$\ell_{max} = \frac{d_1\left(m^{(1)}, m^{(2)}\right)}{\left\|m_k^{(1)}\right\|_{max}} = d_1\left(m^{(1)}, m^{(2)}\right) / \max_k \left(\left(x_k^{(1)}\right)^2 + \left(y_k^{(1)}\right)^2\right)^{1/2}$$

or

$$\ell_2 = \frac{d_1\left(m^{(1)}, m^{(2)}\right)}{\left\|m_k^{(1)}\right\|_{\ell_2}} = d_1\left(m^{(1)}, m^{(2)}\right) / \underset{k}{\text{mean}} \left(\left(x_k^{(1)}\right)^2 + \left(y_k^{(1)}\right)^2\right)^{1/2}.$$

It is interesting that in case of $\ell_{max}$-error, the impact of number of clusters is significant only for $K = 6$. This is due to some outliers that we see in the data which are far from 2 dominating dense regions. Figures 12 and 13 show the results of K-means clustering when $K = 2$ and $K = 6$. Comparison of those figures explains the jump in $\ell_{max}$ for $K = 6$ where one of the clusters contains only 4 points (visually perceived as outliers). This can be a useful procedure for outlier detection – determine the clusters with small percentage of points and define those points as outliers. If the outliers are known, then we can preserve them exactly and encode the remaining normal points by the given Δ-accuracy thus concluding the idea of outlier-Δ-coverage.

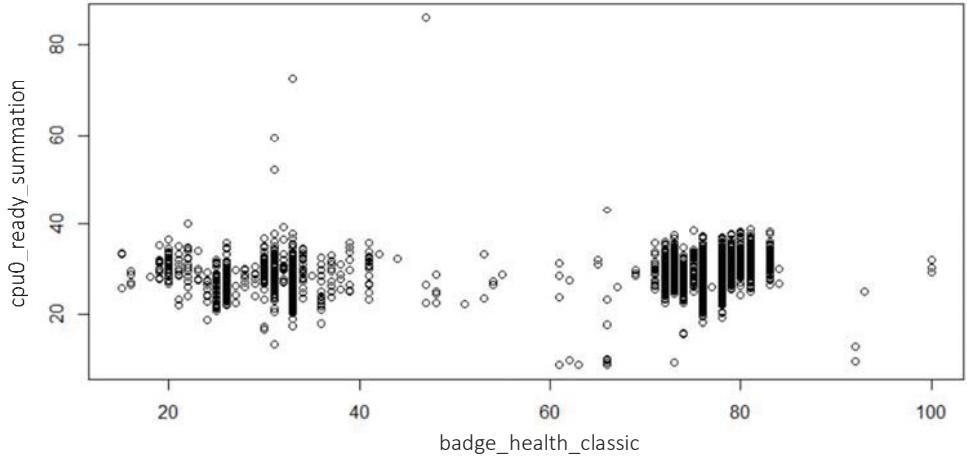

Fig. 10. Original 2D-data that we use for experiments.

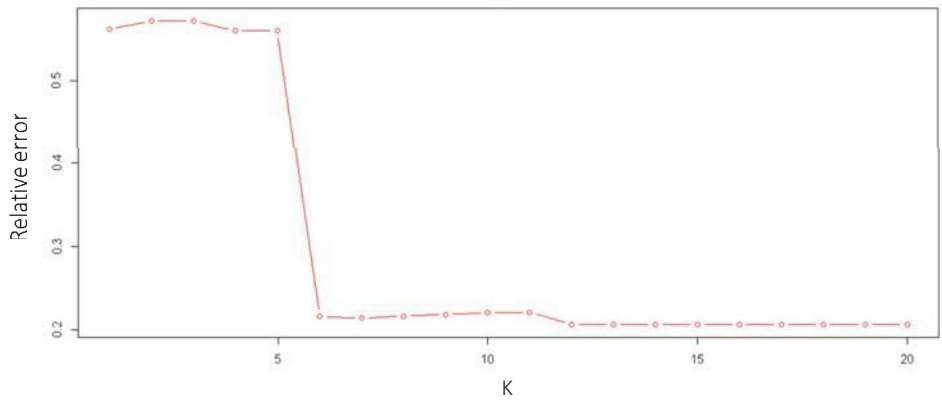

Fig. 11. $\ell_{max}$ error for different number of clusters while applying K-means clustering for optimal coverage.

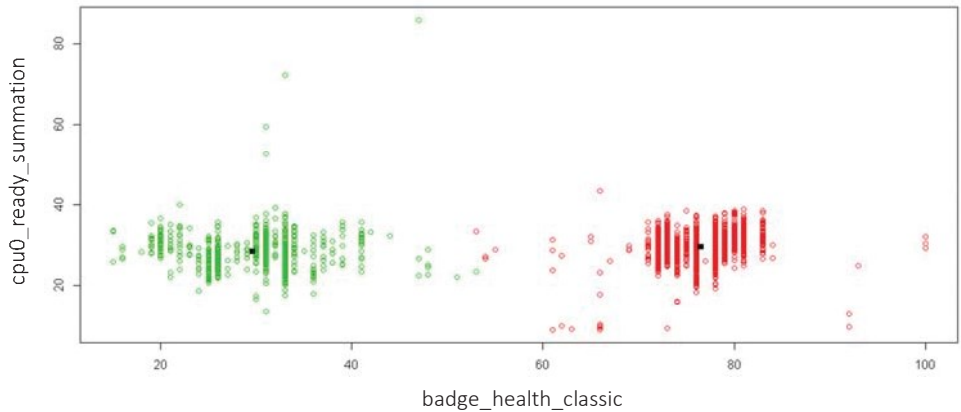

Fig. 12. Clustering of data in Fig. 10 by means of K-means when $K = 2$.

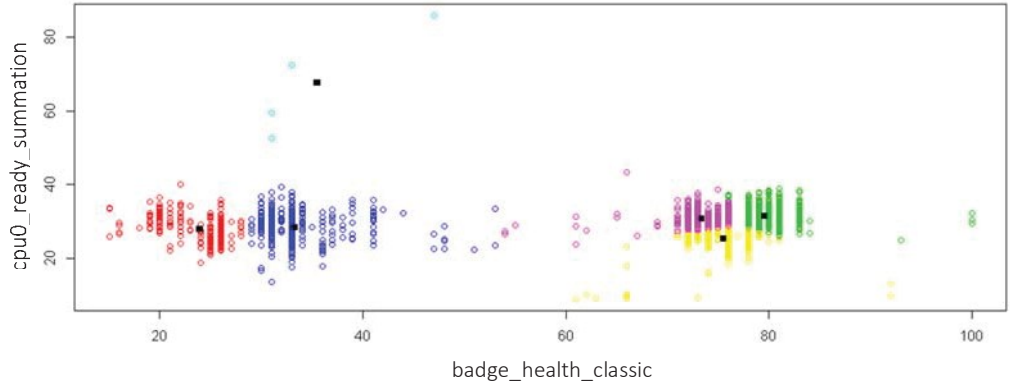

Fig. 13. Clustering of data in Fig. 10 by means of K-means when $K = 6$.

In Figures 14 and 15, we consider another approach for outlier detection by exploring a within-cluster variability and by defining a normal shift from the cluster centroids. For each $j$-th cluster we define a radius $R_j$ of normality from the centroid $c_j = (c_j^1, c_j^2)$ as follows:

$$R_j = 3 * \max_k \left( \left( x_k - c_j^1 \right)^2 + \left( y_k - c_j^2 \right)^2 \right)^{1/2}$$

where the number 3 is a parameter value and the point $(x_k, y_k)$ belongs to the $j$-th cluster. If a point $m_k$ belongs to the $j$-th cluster and its distance from the cluster centroid $c_j$ is less than or equal to $R_j$ then we encode it as $c_j$. If a point lays outside of all normalcy circles, we preserve its exact value without encoding. By changing the number of clusters, we can try to find a minimal (optimal) coverage satisfying the required accuracy.

Fig. 16 shows the values of $\ell_{max}$ for different values of $K$ while constructing the outlier-optimal coverages by the described procedure. Although for small values of $K$ we managed to decrease the errors, for larger values the impact of outliers is still present. Our final recommendation will be combination of both approaches – first, elimination of clusters with small number of points as outliers, and second, calculation of within-cluster variability only for the big ones.

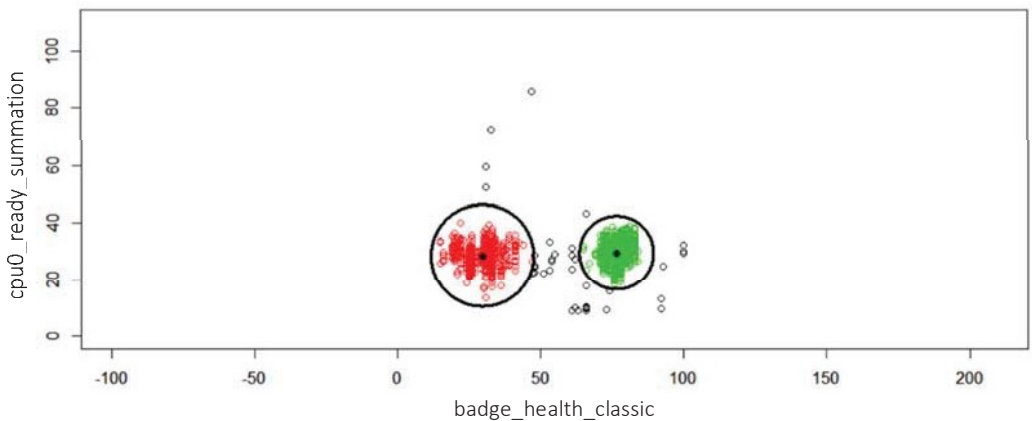

Fig. 14. Result of the $K$-means clustering with $K = 2$ with the corresponding normalcy circles.

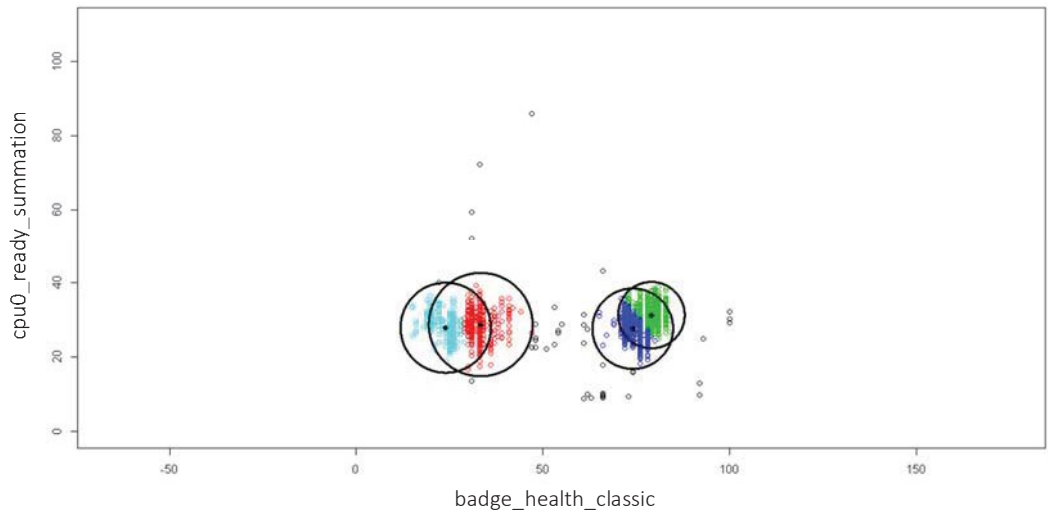

Fig. 15. Result of the $K$-means clustering with $K = 4$ with the corresponding normalcy circles.

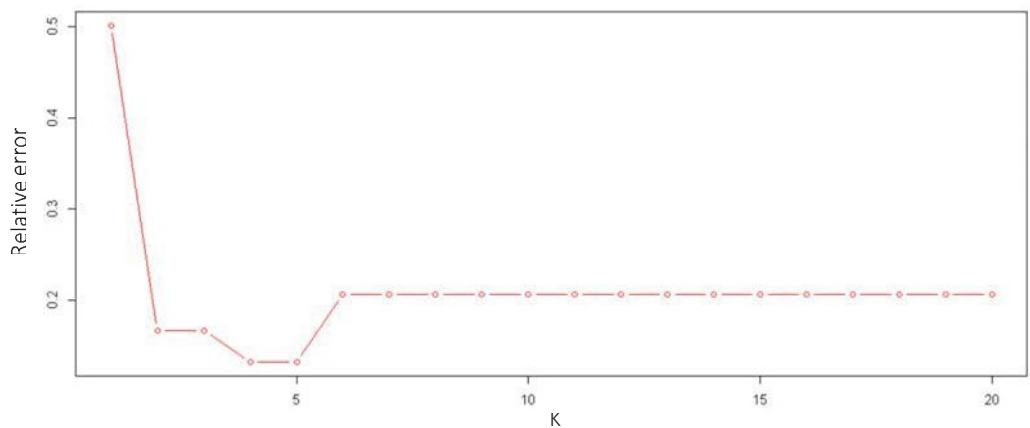

Fig. 16. $\ell_{max}$ -errors for different values of $K$ while constructing the outlier-optimal-coverage by means of $K$-means clustering.

# 2D LDPC Codes for TDMR

Hiroshi Kamabe
Gifu University
1-1, Yanagido, Gifu, 501–1193, Japan
Email: kamabe@ieee.org

Shan Lu
Gifu University
1-1, Yanagido, Gifu, 501–1193, Japan
Email: slu@ieee.org

*Abstract*—Two dimensional magnetic recording systems are investigated extensively to achieve higher recording density. Since the recording density is very high, a small physical defect may cause many two dimensional errors. Hence many two dimensional signal processing of magnetic recording systems are also proposed and investigated. Matcha et al. have proposed a two dimensional LDPC code and investigated the error correcting performance of the code. In this paper we propose a code which is a variant of their code but we can show that the theoretical lower bound of the burst erasure correction capability of our code is better than that of Matcha's code.

## I. Introduction

Since very high speed mobile communication systems, e.g. 5G, and very high resolution broadcast are now available, the demand for digital storage systems with very large capacity increases constantly.

Wood et al. proposed the concept of two dimensional magnetic recording (TDMR) systems to achieve very high recording density [1]. Two dimensional signal processing is one of key techniques of TDMR. The size of a reading head in TDMR is larger than the width of a recording track and there may be no gap between adjacent tracks. Hence a physical defect of a recording medium may cause two dimensional errors.

Matcha et al. have proposed a two dimensional LDPC code and shown a lower bound of the burst erasure correcting capability of the code [2]. By computer simulation they also have shown that the code can correct larger burst erasures than the lower bound.

In this paper we improve their lower bound. Strictly speaking, a code we investigate here is little bit different from the code defined in [2] but the structures of these two codes are almost the same.

## II. Preliminaries

For a finite set $A$, $\#A$ means the number of elements in $A$. The set of integers is denoted by $\mathbb{Z}$. For two positive integers $a$ and $b$, $a \bmod b$ is the remainder when $a$ is divided by $b$, that is, the smallest non negative integer $r$ such that $a = qb + r$ for some integer $q$. We define $\bmod'$ by

$$a \bmod' b = ((a - 1) \bmod b) + 1.$$

We assume that $p$ is an odd prime number throughout this manuscript. For any positive integer $k$, if $k$ is not a multiple of $p$, then we have

$$\begin{aligned} \{ik \bmod p | 1 \le i \le p\} &= \{ik \bmod p | i \in \mathbb{Z}\} \\ &= \{0, 1, \ldots, p - 1\}. \end{aligned}$$

If $(a - b) \bmod p = 0$, then we write $a \equiv b \mod p$. We define a circular permutation $pm(i)$ by

$$pm(i) = (i + 1) \bmod' p.$$

In this paper we index positions of a matrix with positive integers, that is, the set of positions of the matrix of size $n \times m$ is $\{(i, j) \mid 1 \le i \le n, 1 \le j \le m\}$.

Suppose that a parity check matrix $(h_{i,j})$ has the identity matrix of size $3 \times 3$ at the upper left corner of it, that is, the parity check matrix has the following form

$$(h_{ij}) = \begin{pmatrix} 1 & 0 & 0 & \cdots \\ 0 & 1 & 0 & \cdots \\ 0 & 0 & 1 & \cdots \\ \cdots & \cdots & \cdots & \ddots \end{pmatrix}.$$

Consider a code defined by the parity check matrix and suppose that the leftmost three bits of a code word $(c_1, c_2, \ldots, c_n)$ are erased. Since the first parity check equation is

$$c_1 + h_{1,4}c_4 + \cdots h_{1,n}c_n = 0,$$

we can find $c_1$ from the equation and $c_4, c_5, \ldots, c_n$. Similarly, we can find $c_2$ and $c_3$. Matha et al. proposed an LDPC code and shown the burst erasure correcting capability of the code using the principle mentioned above [2].

Next, we consider another parity check matrix whose upper left corner is given as

$$\begin{pmatrix} 1 & 0 & 0 & 0 & 0 & 0 & \cdots \\ 0 & 1 & 0 & 0 & 0 & 0 & \cdots \\ 0 & 0 & 1 & 0 & 0 & 0 & \cdots \\ 1 & 0 & 0 & 1 & 0 & 0 & \cdots \\ 0 & 1 & 1 & 0 & 1 & 0 & \cdots \\ 1 & 1 & 0 & 0 & 0 & 1 & \cdots \\ \cdots & & \cdots & & \cdots & & \ddots \end{pmatrix}$$

Suppose that the leftmost 6 bits of $\boldsymbol{c}$ are erased where $\boldsymbol{c}$ is a code word of a code defined by the second parity check matrix. After the first three bits are recovered from the first 3 parity check equations, the next three bits can also be calculated from the 4th, 5th, and 6th parity check equations with the first three bits. In this paper we use this principle to prove the burst erasure correcting capability of the code we propose.

## III. 2D-LDPC CODE

### A. Code Proposed by Matcha et al.

We explain the construction of the 2 dimensional LDPC code proposed by Matcha et al. [2] as an analogy to the quasi cyclic LDPC code. The code is defined by using a rectangular solid consisting of many (basic) cubes. We explain the structure of the rectangular matrix solid and then the basic cubes.

a

A parity check matrix of a quasi cyclic LDPC code is given as follows:

$$H_{1D} = \begin{pmatrix} I & I & I & \cdots & I \\ R & R^2 & R^3 & \cdots & R^r \\ R^2 & R^4 & R^6 & \cdots & R^{2r} \\ \vdots & \vdots & \vdots & \ddots & \vdots \\ R^{c-1} & R^{2c-2} & R^{3c-3} & \cdots & R^{r(c-1)} \end{pmatrix},$$

where $c$ and $r$ are positive integers and $R$ is a cyclic permutation matrix of size $p \times p$ given as follows:

$$R = \begin{pmatrix} 0 & 1 & 0 & \cdots & 0 \\ 0 & 0 & 1 & \cdots & 0 \\ \vdots & \vdots & \vdots & \ddots & \vdots \\ 0 & 0 & 0 & \cdots & 1 \\ 1 & 0 & 0 & \cdots & 0 \end{pmatrix}.$$

Similarly, Matcha et al. introduced a 2 dimensional LDPC code using permutations of a basic cube $I_d$ of size $p \times p \times p$ defined by

$$I_d(i, j, k) = \begin{cases} 1, & \text{if } i = j = k \\ 0, & \text{otherwise,} \end{cases}$$

[2]. The permutation of $I_d$ will be explained later.

Let $h, w$ and $c$ be positive integers where $c$ is a multiple of $p$. We arrange $hw$ permuted cubes into a rectangular matrix of size $h \times w$ and pile $c$ arranged matrices up, where we consider the size of $I_d$ is $1 \times 1 \times 1$. The size of the resulting rectangular solid is $h \times w \times c$. The position of some basic cube in the rectangular solid is represented as $(I, J, K)$. The position of the cube in the resulting rectangular solid is called the cube position. When we consider the rectangular solid consists of binary bits, the position of a bit in the rectangular solid is called the bit position and represented by $(i, j, k)$. The size of the rectangular solid is $cp \times hp \times wp$ if we measure the size in terms of the bit position.

We define two permutations $P$ and $Q$ of bits in a cube $T$ of size $p \times p \times p$ as follows.

$$(P \circ T)(i, j, k) = T(i, pm(j), k), \quad (1)$$
$$(Q \circ T)(i, j, k) = T(i, j, pm(k)). \quad (2)$$

We define $a(I, J, K)$ and $b_0(I, J, K)$ as follows.

$$a(I, J, K) = I,$$
$$b_0(I, J, K) = \left\lfloor \frac{I-1}{p} \right\rfloor JK. \quad (3)$$

In the rectangular solid defined above, the cube at position $(I, J, K)$ is to be

$$P^{a(I,J,K)} \circ (Q^{b_0(I,J,K)} \circ I_d).$$

The rectangular solid is denoted by $G_0 = (g_0(i, j, k))$ using the bit position notation.

The upper surface of the rectangular solid is a rectangle of size $hp \times pw$ in the bit position. So the rectangular solid can also be constructed by piling $cp$ rectangles of size $hp \times wp$ up. Each rectangle is called a bit layer. Let $c = (c_1, c_2, \ldots, c_{hwp^2})$ be a binary sequence of length $hwp^2$. We define a code as follows: $c$ is a code word if and only if

$$\sum_{\ell=1}^{hwp^2} c_\ell g_0 \left(i, \left\lfloor \frac{\ell-1}{wp} \right\rfloor + 1, \ell \bmod' wp \right) = 0, \quad (4)$$

is satisfied. Hence, in this code every bit layer defines a parity check equation of the code. In the following $c$ is also considered as a plane of size $hp \times wp$. The elements of $c$ are indexed by $(j, k)$ according to the parity check equations.

### B. Our code

We define another code by replacing $b_0$ in (3) of the above definition with the following $b$:

$$b(I, J, K) = \left\lfloor \frac{I-1}{p} \right\rfloor \phi(J) \phi(K), \quad (5)$$

where $\phi$ is defined by

$$\phi(x) = x \bmod' (p-1). \quad (6)$$

A corresponding rectangular solid is denoted by

$$G = (g(i, j, k))$$

using the bit position. Replacing $g_0$ with $g$ in (4), we can define a code $C$ from $G$.

## IV. BURST ERASURE CORRECTION

We examine the burst erasure correcting capability of the code defined in the previous section.

For notational convenience in the rest of this manuscript, the definition of $I_d$ is extended so that

$$I_d(x, y, z) = I_d(x \bmod' p, y \bmod' p, z \bmod' p) \quad (7)$$

for all positive integers $x, y, z$. We note that $I_d$ is periodic with respect to each argument.

We first represent $g(i, j, k)$ using $I_d$. From the definition of our code, we have

$$\begin{aligned} g(i, j, k) &= g(Ip + \bar{i}, Jp + \bar{j}, Kp + \bar{k}) \\ &= I_d(i, j + I + 1, k + \bar{I} \phi(J+1) \phi(K+1)). \\ &= I_d(\bar{i}, \bar{j} + I + 1, \bar{k} + \bar{I} \phi(J+1) \phi(K+1)). \end{aligned} \quad (8)$$

where $\bar{i}, \bar{j}$ and $\bar{k}$ are defined by

$$\bar{i} = i \bmod' p, \quad \bar{j} = j \bmod' p, \quad \bar{k} = k \bmod' p,$$

and $I, J, K$ and $\bar{I}$ are defined so that

$$i = Ip + \bar{i}, \quad j = Jp + \bar{j}, \quad k = Kp + \bar{k}, \quad \bar{I} = \left\lfloor \frac{I}{p} \right\rfloor.$$

When we mention (8), we always use these notation.

The following is our main theorem.

*Theorem 1:* Let $C$ be a code defined in subsection III-B by using a rectangular solid of size $cp \times hp \times wp$ in the bit position where we assume $c \geq p^3 + p^2 + p$. The code $C$ can correct burst erasures of size at least $p \times (2p - 1)$.

The code rate of our code is

$$\frac{hwp - c}{hwp}$$

and is upper bounded by

$$\frac{hwp - (p^3 + p^2 + p))}{hwp}.$$

By increasing $h$ and $w$ we can obtain better code rates but the lower bound of the size of correctable burst errors does not change if $p$ does not change. When $hw$ tends to $\infty$, the upper bound of the code rate tends to 1.

We need the following lemmas to prove Theorem 1.

*Lemma 2:*

1) for all positive integers $j, k$ and for all non-negative integers $m$, we have

$$\#\{i : g(mp + i, j, k) = 1, 1 \leq i \leq p\} \leq 1. \quad (9)$$

2) for all positive integers $i, k$ and for all non-negative integers $m$, we have

$$\#\{j : g(i, mp + j, k) = 1, 1 \leq j \leq p\} \leq 1. \quad (10)$$

3) for all positive integers $i, j$ and for all non-negative integers $m$, we have

$$\#\{k : g(i, j, mp + k) = 1, 1 \leq k \leq p\} \leq 1. \quad (11)$$

*Lemma 3:* For any positive integers $j$ and $k$, there exists an $i$ with

$$g(i, j, k) = 1 \text{ and } i \leq p^2 + p.$$

*Lemma 4:* Let $j$ be positive integer and let $k_0$ be a non negative multiple of $p$. For every $k'$ with $1 \leq k' \leq p - 1$, we can find an $i$ which satisfies the following conditions:

(A) $g(i, j, k_0 + k') = 1$;
(B) for all $\tilde{k}$ with $0 \leq \tilde{k} < 2p$ and $k' \neq \tilde{k}$,

$$g(i, j, k_0 + \tilde{k}) = 0;$$

(C) $g(i, j, k_0 + 2p) = 1$;
(D) $i \leq p^3 + p^2 + p$.

We have the following technical lemma.

*Lemma 5:* Let $A, B$ be integers in $\{1, \ldots, p-1\}$ with $A \neq B$. Let $C, s \in \{0, 1, \ldots, p-1\}$. Then there is an $I$ which satisfies the following simultaneously,

$$\begin{aligned}
k_1, k_2 &\in \{0, 1, \ldots, p-1\}, \\
k_1 &\equiv C + IA \mod p, \\
k_2 &\equiv C + IB \mod p, \\
k_1 &\leq s \leq k_2.
\end{aligned}$$

*Lemma 6:* Let $j$ be positive integers. Let $k_0$ be a nonnegative multiple of $p$. Suppose that $m$ and $s$ are integers with $1 \leq m \leq p$ and $1 \leq s \leq p - 1$. Then there is an integer $i$ which satisfies the following;

$$g(i, j, k_0 + p + m) = 1, \quad (12)$$

for some $k_1$ with $0 \leq k_1 \leq s$

$$g(i, j, k_0 + k_1) = 1, \quad (13)$$

for some $k_2$ with $2p + s \leq k_2 \leq 3p - 1$

$$g(i, j, k_0 + k_2) = 1, \quad (14)$$

$$i \leq p^2 + p^2 + p. \quad (15)$$

*Lemma 7:* Let $j$ be a positive integer and let $k_0$ be a nonnegative multiple of $p$. For every pair $(k_1, k_2)$ with $1 \leq k_1, k_2 \leq p$, there is an $i$ such that

$$g(i, j, k_0 + k_1) = g(i, j, k_0 + 2p + k_2) = 1, \quad \text{and} \quad i \leq p^2 + p.$$

*Lemma 8:* Suppose that $g(i, j, k) = 1$. Then we have $g(i, j', k') = 0$ for all $(j', k')$ with $0 < |j - j'| < p$, $0 < j'$ and $0 < k$.

Now we prove Theorem 1.

## V. Conclusion

We have proposed a 2 dimensional LDPC code which is a variant of a code proposed by Matcha et al. [2] and investigated the burst erasure correcting capability of our code. We should evaluate the error correcting performance by computer simulation with the generalized belief propagation.

### Acknowledgment

This work was partially supported by KAKENHI 16K14267(C), 19K11822(C), 18H01133(B) and ASRC (Advanced Storage Research Consortium). <

# A Universal Source Code Based on the AIVF Coding Strategy

Hirosuke Yamamoto
The University of Tokyo
Kashiwa-shi, Chiba, 277-0871, Japan.
hirosuke@ieee.org

Kengo Hashimoto
University of Fukui
Fukui-shi, Fukui, 910-8507, Japan.
kengo3.1415@gmail.com

Kenichi Iwata
University of Fukui
Fukui-shi, Fukui, 910-8507, Japan.
k-iwata@u-fukui.ac.jp

## Summary

We propose a new universal source code based on the coding strategy of AIVF (almost instantaneous variable-to-fixed) code [1]. The proposed code can also be considered as an extended code of the LZW code (Welch code) [2] and the Yokoo code [3].

The LZW code and Yokoo code are classified into the family of the LZ78 code [4], which uses the so-called incremental parsing or similar parsing, and they can be implemented by using a parse tree. In the LZW coding, serial numbers are assigned to all nodes including leaves, complete internal node, and incomplete internal nodes in a parse tree. A source sequence is parsed by the longest match in the parse tree, and the parsed string is encoded by the number assigned to the node of the longest match. The incremental-like parsing is realized by growing one child from the longest match node in the parse tree after each encoding of a parsed string. But, since a number is assigned to a node even after the node becomes a complete node, which cannot become the longest match node, the LZW has a coding loss.

On the other hand, Yokoo code is designed for binary sources and uses a full binary tree as a parse tree. Serial numbers are assigned to only leaves, and the parse tree is grown to realize the incremental parsing by generating all children at once from the longest match leaf. But, Yokoo code cannot attain good compression rate for non-binary sources because the number of leaves becomes very large as the encoding progresses, and many serial numbers of leaves are wasted without using. Recently, in order to overcome the defect of Yokoo code for non-binary sources, Arimura [5] proposed two-path coding scheme such that all leaves of a parse tree are generated for a given source sequence and unnecessary leaves are trimmed in the first path, and then the source sequence is encoded by using the trimmed parse tree in the second path. But, Arimura code requires two paths in encoding and large memory space in the first path. Furthermore, the information of trimmed parse tree must be encoded and included in the prefix of the codeword sequence.

We also note that the above codes have another defect such that successive parsed strings are encoded independently although they are generally dependent.

Our universal code uses the same coding strategy as the

AIVF codes [1], which is not a universal code. The AIFV code can attain better compression rate than Tunstall code by using fixed multiple parse trees, and the multiple parse trees can be integrated into a single parse tree [6]. In our universal code, the parse tree grows by generating one child from the longest match node in the same way as the LZW code. However, serial numbers are assigned to only leaves and incomplete internal nodes in the same way as the AIVF code. Furthermore, if a parsed string is encoded at an incomplete internal node, then the next parsed string is efficiently encoded by removing some parts of the parsing tree similarly to the AIVF coding.

We prove theoretically that the proposed universal code is asymptotically optimal for any stationary ergodic sources in the same way as the LZ78 code and LZW code, and we also show by comparing the compression rates for several corpora that the proposed universal code can attain better compression rate than the LZW code practically. It is worth noting that the coding strategy of the proposed universal code can be applied to other variants of LZ78 codes, e.g. Yokoo's variant of LZW code [7], to improve their compression rates.

# Modern applications of convolutional codes

Ángela Barbero and Øyvind Ytrehus

May 24, 2019

## Abstract

During the first decades of coding and information theory, binary convolutional codes were popular in practical applications due to simple implementations of encoders and maximum likelihood decoders. Recent research has produced new results on the construction of nonbinary convolutional codes [1], mainly over finite extension fields of characteristic 2. Although these codes are more powerful than their binary relatives, they are less suited for the traditional applications in communication, since the complexity of decoding is prohibitive for all but the simplest codes. However, these codes offer great performance over erasure channels, as well as fast error detection on channels with errors. We describe some relevant applications, including protocols for streaming of IP packets over the Internet, [2], codes for dynamic distributed storage [3][4], and codes for fast error detection [5].

# On Overflow Probability of Variable-to-Fixed Length Codes for Non-Stationary Sources

Shigeaki Kuzuoka

Faculty of Systems Engineering, Wakayama University

Email: kuzuoka@ieee.org

## Extended Abstract

Let $\boldsymbol{X} = \{X^n\}_{n=1}^{\infty}$ be a *general source* [1] with the source alphabet $\mathcal{X}$; i.e., $\boldsymbol{X}$ is a sequence of random variables $X^n$ on the $n$-fold Cartesian product $\mathcal{X}^n$. In this paper, we assume the *consistency* of the source; i.e., any $x^n = (x_1, x_2, \dots, x_n) \in \mathcal{X}^n$ satisfies

$$P_{X^n}(x^n) = \sum_{\hat{x} \in \mathcal{X}} P_{X^{n+1}}(x^n \circ \hat{x})$$

where $\circ$ means the concatenation. Further we assume that $\mathcal{X}$ is finite. However we do not assume the stationarity or ergodicity of the source.

A *variable-to-fixed length code* (VF code) is characterized by a set of finite strings (words)

$$\mathcal{C}_n \subseteq \bigcup_{\ell=1}^{\infty} \mathcal{X}^\ell$$

satisfying the following conditions: (i) the cardinality $|\mathcal{C}_n|$ is at most $2^n$ and (ii) $\mathcal{C}_n$ is proper and complete; i.e., any infinite string $x_1 x_2 \cdots \in \mathcal{X}^\infty$ has one and only one prefix $x^\ell \in \mathcal{C}_n$. Note that any word in $\mathcal{C}_n$ can be represented (encoded) losslessly by using $n$-bits. The *empirical compression rate* (ECR) associated with a word $x^\ell \in \mathcal{C}_n$ of length $\ell$ is defined as $n/\ell$.

In this study, we investigate the probability that the ECR exceeds a given threshold $R$; i.e., the probability that $n/\ell > R$ (or equivalently $\ell \leq \lceil n/R \rceil - 1$) holds. More precisely we consider the probability $e(\mathcal{C}_n, R | \boldsymbol{X})$ defined as

$$e(\mathcal{C}_n, R | \boldsymbol{X}) \triangleq \sum_{\ell=1}^{\lceil n/R \rceil - 1} \sum_{x^\ell \in \mathcal{C}_n \cap \mathcal{X}^\ell} P_{X^\ell}(x^\ell).$$

A threshold $R$ is said be *achievable* if there exists a sequence $\{\mathcal{C}_n\}_{n=1}^{\infty}$ of VF codes satisfying

$$\lim_{n \to \infty} e(\mathcal{C}_n, R | \boldsymbol{X}) = 0.$$

The infimum of achievable thresholds is denoted by $\rho(\boldsymbol{X})$. Then we have the following result.

**Theorem 1.**

$$\rho(\boldsymbol{X}) = \overline{H}(\boldsymbol{X})$$

where $\overline{H}(\boldsymbol{X})$ denotes the spectral sup-entropy rate of $\boldsymbol{X}$ [1]:

$$\overline{H}(\boldsymbol{X}) \triangleq \text{p-}\limsup_{n \to \infty} \frac{1}{n} \log \frac{1}{P_{X^n}(X^n)}.$$

**Remark 1** (Related work). The overflow probability and the large deviation performance of VF codes for finite-state sources were investigated by Merhav and Neuhoff [2]. The competitive optimality of VF codes for i.i.d. sources were studied by Yamamoto and Yokoo [3]. Furthermore, several performance criteria of VF codes were introduced and studied by Arimura [4].

**Remark 2.** It is not hard to generalize Theorem 1 to the case of $\varepsilon$-coding; i.e., the case where $\limsup e(\mathcal{C}_n, R | \boldsymbol{X}) \leq \varepsilon$ is required.

*Proof Sketch of Theorem 1:*

(Converse part) Given a VF code $\mathcal{C}_n$, we can construct a Fixed-to-Fixed length code from $\mathcal{X}^{\lceil n/R \rceil}$ to $n$-bits codewords with the decoding error probability $e(\mathcal{C}_n, R | \boldsymbol{X})$. Hence, from the converse coding theorem for FF codes [1, Theorem 1.3.1], we have $\rho(\boldsymbol{X}) \geq \overline{H}(\boldsymbol{X})$.

(Direct part) For any infinite string $\boldsymbol{x} = x_1 x_2 \cdots \in \mathcal{X}^\infty$, let $\ell = \ell(\boldsymbol{x})$ be the smallest integer such that

$$\log \frac{1}{P_{X^\ell}(x^\ell)} + \ell \gamma_n > n$$

where $\gamma_n \triangleq (\log n)/\sqrt{n}$. Then let $\boldsymbol{c}(\boldsymbol{x})$ be the prefix $(x_1, \dots, x_{\ell(\boldsymbol{x})})$ of $\boldsymbol{x}$ for each $\boldsymbol{x} \in \mathcal{X}^\infty$, and let us define the code as $\mathcal{C}_n^* \triangleq \{\boldsymbol{c}(\boldsymbol{x}) : \boldsymbol{x} \in \mathcal{X}^\infty\}$. We can show that $\mathcal{C}_n^*$ satisfies $|\mathcal{C}_n^*| \leq 2^n$ for sufficiently large $n$, and

$$e(\mathcal{C}_n^*, R | \boldsymbol{X}) \leq \Pr\left\{ \frac{1}{\hat{n}} \log \frac{1}{P_{X^{\hat{n}}}(X^{\hat{n}})} + \gamma_n > R \right\}$$

where $\hat{n} \triangleq \lceil n/R \rceil - 1$. Hence, from the definition of $\overline{H}(\boldsymbol{X})$, we have $\rho(\boldsymbol{X}) \leq \overline{H}(\boldsymbol{X})$. ∎

## Acknowledgment

The work was supported in part by JSPS KAKENHI Grant Number 18K04141.

# On Channel Coding with Cost Constraint on Delivery

Mikihiko Nishiara

Shinshu University

4-17-1 Wakasato, Nagano-city, Nagano, 380-8553 Japan

Email: mikihiko@shinshu-u.ac.jp

*Abstract*—In the source coding problem with cost constraint, a cost function is defined over the code alphabet. This can be regarded as a noiseless channel coding problem with cost constraint. In this case, we will not distinguish between the input alphabet and the output alphabet of the channel. However, we must distinguish them for a noisy channel. In the channel coding problem with cost constraint so far, the cost function is defined over the input alphabet of the noisy channel. In this paper, we define the cost function over the output alphabet of the channel. And, the cost is paid only after the received word is observed. Note that the cost is a random variable even if the codeword is fixed. We show the channel capacity with cost constraint defined over the output alphabet. Moreover, we generalize it to tolerate a small decoding error and a small cost overrun.

## I. INTRODUCTION

In the source coding problem with cost constraint, a cost function is defined over the code alphabet. This can be regarded as a noiseless channel coding problem with cost constraint [1]. In this situation, we will not distinguish between the input alphabet and the output alphabet of the channel. However, for a noisy channel, the output alphabet is generally distinguished from the input alphabet. We then have two options as the domain of the cost function.

So far, the cost function has been defined over the input alphabet of the noisy channel in the channel coding problem with cost constraint [2], [3], [4]. In this paper, we define the cost function over the output alphabet of the channel. And, the cost is paid only after the received word is observed because the encoder cannot control the exact cost at the decoder. Technically, the cost is a random variable even if the codeword is fixed. We show the channel capacity with cost constraint defined over the output alphabet. Moreover, we generalize it to tolerate a small decoding error and a small cost overrun.

The idea of the cost paid on delivery came from a study in the paper [5] entitled "bits through queues." In that paper, a single server queueing system is regarded as a channel, and the time intervals of packets convey the information. One of the features of that system is that the length of received word differs from the length of codeword transmitted by the encoder. Therefore we can consider the length of received word as a stochastic cost. In that paper, the situation is formulated as an ordinary channel coding problem dealing with information per unit time. On the other hand, with the scheme of this paper, the same situation is also formulated as a channel coding problem with cost constraint dealing with information per symbol and cost per symbol.

## II. COST PAID ON DELIVERY

### A. Preliminary

Let $\mathcal{X}$ and $\mathcal{Y}$ be two finite sets. Consider general channel $\boldsymbol{W} \triangleq \{W^n\}_{n=1}^{\infty}$ with $\mathcal{X}$ and $\mathcal{Y}$ as input and output alphabet, respectively. This means that $W^n(\cdot|\boldsymbol{x}), \boldsymbol{x} \in \mathcal{X}^n$ is a distribution over $\mathcal{Y}^n$. We define a real-valued function $c_n$ on the output alphabet $\mathcal{Y}^n$ and call it the cost function. For $\boldsymbol{y} \in \mathcal{Y}^n$, $c_n(\boldsymbol{y})$ is called the cost of $\boldsymbol{y}$.

We want to inform the destination of one of $M_n$ messages through this channel. Let $\mathcal{M}_n \triangleq \{1, \ldots, M_n\}$ denote the set of messages. A selected message $m \in \mathcal{M}_n$ is encoded by the encoder $\varphi_n : \mathcal{M}_n \to \mathcal{X}^n$ into codeword $\varphi_n(m)$, which is fed into the channel. Observing the output of the channel, the decoder guesses the message selected at the encoder side. The guess of the decoder is not necessarily correct. The probability that the guess differs from the selected message is called the error probability and represented by

$$\varepsilon_n \triangleq 1 - \frac{1}{M_n} \sum_{m \in \mathcal{M}_n} W^n(\psi_n^{-1}(m)|\varphi_n(m)), \qquad (1)$$

where $\psi_n^{-1}(m) \triangleq \{\boldsymbol{y} \in \mathcal{Y}^n | \psi_n(\boldsymbol{y}) = m\}$, which is called the decoding region of $m$ or $\varphi_n(m)$. A pair $(\varphi_n, \psi_n)$ of encoder $\varphi_n$ and decoder $\psi_n$ is called a code.

Every logarithm in this paper is the natural logarithm. Generally, the distribution of a random variable $Z$ is denoted by $P_Z$. For a sequence of real-valued random variables $\{Z_n\}_{n=1}^{\infty}$, we define [4]

$$\text{p-}\limsup_{n\to\infty} Z_n \triangleq \inf\left\{\theta \, \Big| \, \lim_{n\to\infty} \Pr\{Z_n > \theta\} = 0\right\}, \qquad (2)$$

$$\text{p-}\liminf_{n\to\infty} Z_n \triangleq \sup\left\{\theta \, \Big| \, \lim_{n\to\infty} \Pr\{Z_n < \theta\} = 0\right\}. \qquad (3)$$

Let $\boldsymbol{X} \triangleq \{X^n\}_{n=1}^{\infty}$ and $\boldsymbol{Y} \triangleq \{Y^n\}_{n=1}^{\infty}$. Generally, given that $\boldsymbol{X}$ is fed into the channel, the output is denoted by $\boldsymbol{Y}(\boldsymbol{X})$. For each $n$, given that $X^n$ is fed into the channel, the output is denoted by $Y^n(X^n)$. That is, we define

$$P_{Y^n(X^n)}(\boldsymbol{y}) \triangleq \sum_{\boldsymbol{x} \in \mathcal{X}^n} P_{X^n}(\boldsymbol{x}) W^n(\boldsymbol{y}|\boldsymbol{x}), \quad \boldsymbol{y} \in \mathcal{Y}^n. \qquad (4)$$

Let $U_M$ denote the random variable uniformly distributed over the set of natural integers not exceeding $M$.

## B. Achievability and Capacity

*Definition 1:* Rate $R$ is said to be $\Gamma$-achievable if and only if there exists a sequence of codes $\{(\varphi_n, \psi_n)\}_{n=1}^{\infty}$ satisfying

$$\lim_{n \to \infty} \varepsilon_n = 0, \tag{5}$$

$$\liminf_{n \to \infty} \frac{1}{n} \log M_n \geq R, \tag{6}$$

$$\text{p-}\limsup_{n \to \infty} \frac{1}{n} c_n(Y^n) \leq \Gamma, \tag{7}$$

where $Y^n = Y^n(\varphi_n(U_M))$ on (7).

*Definition 2:* For $\Gamma$, we define

$$C(\Gamma) \triangleq \sup\{R | R \text{ is } \Gamma\text{-achievable}\} \tag{8}$$

and call it $\Gamma$-channel capacity.

*Definition 3:* For input $\boldsymbol{X}$ and output $\boldsymbol{Y} = \boldsymbol{Y}(\boldsymbol{X})$ of the channel, we define

$$\underline{I}(\boldsymbol{X}; \boldsymbol{Y}) = \text{p-}\liminf_{n \to \infty} \frac{1}{n} \log \frac{W^n(Y^n|X^n)}{P_{Y^n}(Y^n)}. \tag{9}$$

*Definition 4:* For arbitrary output $\boldsymbol{Y}$ of the channel, we define

$$\overline{c}(\boldsymbol{Y}) \triangleq \text{p-}\limsup_{n \to \infty} \frac{1}{n} c_n(Y^n). \tag{10}$$

## C. Fundamental Theorem

We have the following theorem.

*Theorem 1:* $\Gamma$-capacity is given by

$$C(\Gamma) = \sup_{\boldsymbol{X}: \overline{c}(\boldsymbol{Y}) \leq \Gamma} \underline{I}(\boldsymbol{X}; \boldsymbol{Y}), \tag{11}$$

where $\boldsymbol{Y} = \boldsymbol{Y}(\boldsymbol{X})$.

To prove the direct part of Theorem 1, we prepare the following lemma, which is a slightly extended version of Feinstein's Lemma [4, Lemma 3.4.1].

*Lemma 1:* Let us consider an arbitrary random variable $X^n$ on $\mathcal{X}^n$ and arbitrary values $R > 0$ and $\gamma > 0$. Let $Y^n = Y^n(X^n)$. Given an arbitrary subset $\mathcal{S}_n \subset \mathcal{Y}^n$, there exists a code $(\varphi_n, \psi_n)$ satisfying

$$\frac{1}{n} \log M_n \geq R, \tag{12}$$

$$\varepsilon_n \leq \lambda_n, \tag{13}$$

$$W^n(\mathcal{S}_n | \varphi_n(m)) > 1 - \lambda_n, \quad m \in \mathcal{M}_n, \tag{14}$$

where $\mathcal{M}_n \triangleq \{1, \dots, M_n\}$ denotes the message set and

$$\lambda_n \triangleq \Pr\left\{ \frac{1}{n} \log \frac{W^n(Y^n|X^n)}{P_{Y^n}(Y^n)} < R + \gamma \right\}$$
$$+ \Pr\{Y^n \notin \mathcal{S}_n\} + e^{-n\gamma}. \tag{15}$$

*Proof:* Omitted. $\qquad \Box$

*Proof of Theorem 1:* We give the proof of the direct part of Theorem 1, while the converse part is omitted because it is proved by following the same lines as the original case [4, Theorem 3.6.1]

Let

$$R_0 \triangleq \sup_{\boldsymbol{X}: \overline{c}(\boldsymbol{Y}) \leq \Gamma} \underline{I}(\boldsymbol{X}; \boldsymbol{Y}) \tag{16}$$

and consider arbitrary $R < R_0$. For some $\gamma > 0$, there exists an $\boldsymbol{X}$ such that

$$R + \gamma < \underline{I}(\boldsymbol{X}; \boldsymbol{Y}), \tag{17}$$

$$\overline{c}(\boldsymbol{Y}) \leq \Gamma, \tag{18}$$

where $\boldsymbol{Y} = \boldsymbol{Y}(\boldsymbol{X})$. From the definition of $\overline{c}(\boldsymbol{Y})$, we have

$$\lim_{n \to \infty} \Pr\left\{ \frac{1}{n} c_n(Y^n) > \Gamma + \frac{1}{k} \right\} = 0 \tag{19}$$

for any natural number $k$. Moreover, for all $n$ large enough,

$$\Pr\left\{ \frac{1}{n} c_n(Y^n) > \Gamma + \frac{1}{k} \right\} < \frac{1}{k}. \tag{20}$$

With the diagonal line argument, we can show that there exists a vanishing series $\gamma_n \to 0$ satisfying

$$\Pr\left\{ \frac{1}{n} c_n(Y^n) > \Gamma + \gamma_n \right\} < \gamma_n. \tag{21}$$

We define

$$\mathcal{S}_n \triangleq \left\{ \boldsymbol{y} \in \mathcal{Y}^n \left| \frac{1}{n} c_n(\boldsymbol{y}) \leq \Gamma + \gamma_n \right. \right\} \tag{22}$$

and apply Lemma 1. Then, we have a code $(\varphi_n, \psi_n)$ satisfying (12), (13), and (14). The first and second terms of the right hand side of (15) vanish due to (17) and (21), respectively. Therefore, we have $\lambda_n \to 0$. This means $\varepsilon_n \to 0$.

Furthermore, (14) yields

$$\Pr\left\{ \frac{1}{n} c_n(Y^n(\varphi_n(U_{M_n}))) > \Gamma + \gamma_n \right\} \tag{23}$$

$$= \frac{1}{M_n} \sum_{m \in \mathcal{M}_n} \Pr\left\{ \frac{1}{n} c_n(Y^n(\varphi_n(m))) > \Gamma + \gamma_n \right\} \tag{24}$$

$$= 1 - \frac{1}{M_n} \sum_{m \in \mathcal{M}_n} W^n(\mathcal{S}_n | \varphi_n(m)) < \lambda_n \to 0. \tag{25}$$

Then, we have

$$\Gamma \geq \text{p-}\limsup_{n \to \infty} \left( \frac{1}{n} c_n(Y^n(\varphi_n(U_{M_n}))) - \gamma_n \right) \tag{26}$$

$$= \text{p-}\limsup_{n \to \infty} \frac{1}{n} c_n(Y^n(\varphi_n(U_{M_n}))). \tag{27}$$

Hence, we can conclude that rate $R$ is $\Gamma$-achievable and that $C(\Gamma) \geq R_0$. $\qquad \Box$

## III. ALLOWING SMALL DECODING ERROR

We weaken the requirement for the error probability.

*Definition 5:* Rate $R$ is said to be $(\varepsilon, \Gamma)$-achievable if and only if there exists a sequence of codes $\{(\varphi_n, \psi_n)\}_{n=1}^{\infty}$ satisfying

$$\lim_{n \to \infty} \varepsilon_n \leq \varepsilon, \tag{28}$$

$$\liminf_{n \to \infty} \frac{1}{n} \log M_n > R, \tag{29}$$

$$\overline{c}(\{Y^n(\varphi_n(U_{M_n}))\}_{n=1}^{\infty}) \leq \Gamma. \tag{30}$$

*Definition 6:* For $\varepsilon$, $\Gamma$, we define

$$C(\varepsilon,\Gamma) \triangleq \sup\{R | R \text{ is } (\varepsilon,\Gamma)\text{-achievable}\} \quad (31)$$

and call it $(\varepsilon,\Gamma)$-channel capacity.

In order to find a formula for $(\varepsilon,\Gamma)$-channel capacity, we introduce the following quantity.

*Definition 7:* For an input $\boldsymbol{X}$ and an output $\boldsymbol{Y} = \boldsymbol{Y}(\boldsymbol{X})$ of the channel, we define

$$\underline{I}_\varepsilon(\boldsymbol{X};\boldsymbol{Y}) \triangleq \sup\Bigg\{R \,\Bigg|$$

$$\limsup_{n\to\infty} \Pr\left\{\frac{1}{n}\log\frac{W^n(Y^n|X^n)}{P_{Y^n}(Y^n)} < R\right\} \le \varepsilon\Bigg\}. \quad (32)$$

Then, we have the following theorem.

*Theorem 2:* $(\varepsilon,\Gamma)$-channel capacity is given by

$$C(\varepsilon,\Gamma) = \sup_{\boldsymbol{X}:\bar{c}(\boldsymbol{Y})<\Gamma} \underline{I}_\varepsilon(\boldsymbol{X};\boldsymbol{Y}), \quad (33)$$

where $\boldsymbol{Y} = \boldsymbol{Y}(\boldsymbol{X})$.

To prove the direct part of Theorem 2, we apply Lemma 2 given as follows instead of Lemma 1.

*Lemma 2:* Let us consider an arbitrary random variable $X^n$ on $\mathcal{X}^n$ and arbitrary values $R > 0$ and $\gamma > 0$. Let $Y^n = Y^n(X^n)$. Given an arbitrary subset $\mathcal{S}_n \subset \mathcal{X}^n$, there exists a code $(\varphi_n, \psi_n)$ whose codewords are in $\mathcal{S}_n$, and it satisfies

$$\frac{1}{n}\log M_n \ge R, \quad (34)$$

$$\varepsilon_n \le \Pr\left\{\frac{1}{n}\log\frac{W^n(Y^n|X^n)}{P_{Y^n}(Y^n)} < R+\gamma\right\}$$
$$+ \Pr\{X^n \notin \mathcal{S}_n\} + e^{-n\gamma}. \quad (35)$$

*Proof:* First of all, let

$$\mathcal{B}_n(\boldsymbol{x}) \triangleq \left\{\boldsymbol{y} \in \mathcal{Y}^n \,\bigg|\, \frac{1}{n}\log\frac{W^n(\boldsymbol{y}|\boldsymbol{x})}{P_{Y^n}(\boldsymbol{y})} \ge R+\gamma\right\}, \quad (36)$$

$$\lambda_n \triangleq \Pr\{Y^n \notin \mathcal{B}_n(X^n)\} + \Pr\{X^n \notin \mathcal{S}_n\} + e^{-n\gamma}. \quad (37)$$

We iteratively choose codewords among $\mathcal{S}_n$. As the first codeword $\varphi_n(1)$, we arbitrarily choose an $\boldsymbol{x} \in \mathcal{S}_n$ satisfying

$$W^n(\mathcal{B}_n(\boldsymbol{x})|\boldsymbol{x}) \ge 1 - \lambda_n, \quad (38)$$

and define the decoding region for $\varphi_n(1)$ as $\psi_n^{-1}(1) \triangleq \mathcal{B}_n(\varphi_n(1))$. Thereafter, as the $m$th codeword $\varphi_n(m)$, we arbitrarily choose an $\boldsymbol{x} \in \mathcal{S}_n$ satisfying

$$W^n\left(\mathcal{B}_n(\boldsymbol{x}) \setminus \bigcup_{m' < m} \psi_n^{-1}(m')\,\bigg|\, \boldsymbol{x}\right) \ge 1 - \lambda_n, \quad (39)$$

and define the decoding region for $\varphi_n(m)$ as

$$\psi_n^{-1}(m) \triangleq \mathcal{B}_n(\varphi_n(m)) \setminus \bigcup_{m' < m} \psi_n^{-1}(m'). \quad (40)$$

Assume that we have taken $M_n$ codewords and cannot choose any more codeword. Here, let $\mathcal{D} \triangleq \bigcup_{m \le M_n} \psi_n^{-1}(m)$.

To evaluate the coding rate of the obtained code, we bound

$$\Pr\{Y^n \in \mathcal{B}_n(X^n)\}$$
$$= \Pr\{Y^n \in \mathcal{B}_n(X^n) \cap \mathcal{D}\}$$
$$+ \Pr\{Y^n \in \mathcal{B}_n(X^n) \setminus \mathcal{D}, X^n \in \mathcal{S}_n\}$$
$$+ \Pr\{Y^n \in \mathcal{B}_n(X^n) \setminus \mathcal{D}, X^n \notin \mathcal{S}_n\} \quad (41)$$

term by term. The first term is bounded by

$$\Pr\{Y^n \in \mathcal{B}_n(X^n) \cap \mathcal{D}\} \quad (42)$$
$$\le \Pr\{Y^n \in \mathcal{D}\} \quad (43)$$
$$= \sum_{m \le M_n} P_{Y^n}(\psi_n^{-1}(m)) \quad (44)$$
$$\le \sum_{m \le M_n} P_{Y^n}(\mathcal{B}_n(\varphi_n(m))) \quad (45)$$
$$= \sum_{m \le M_n} \sum_{\boldsymbol{y} \in \mathcal{B}_n(\varphi_n(m))} P_{Y^n}(\boldsymbol{y}) \quad (46)$$
$$\le \sum_{m \le M_n} \sum_{\boldsymbol{y} \in \mathcal{B}_n(\varphi_n(m))} W^n(\boldsymbol{y}|\varphi_n(m))e^{-n(R+\gamma)} \quad (47)$$
$$\le M_n e^{-n(R+\gamma)}. \quad (48)$$

The second term is bounded by

$$\Pr\{Y^n \in \mathcal{B}_n(X^n) \setminus \mathcal{D}, X^n \in \mathcal{S}_n\} \quad (49)$$
$$= \sum_{\boldsymbol{x} \in \mathcal{S}_n} P_{X^n}(\boldsymbol{x})W^n(\mathcal{B}_n(\boldsymbol{x}) \setminus \mathcal{D}|\boldsymbol{x}) \quad (50)$$
$$< \sum_{\boldsymbol{x} \in \mathcal{S}_n} P_{X^n}(\boldsymbol{x})(1-\lambda_n) \le 1 - \lambda_n. \quad (51)$$

And the third term is bounded by

$$\Pr\{Y^n \in \mathcal{B}_n(X^n) \setminus \mathcal{D}, X^n \notin \mathcal{S}_n\} \le \Pr\{X^n \notin \mathcal{S}_n\}. \quad (52)$$

Therefore, it follows that

$$\Pr\{Y^n \in \mathcal{B}_n(X^n)\}$$
$$\le M_n e^{-n(R+\gamma)} + (1-\lambda_n) + \Pr\{X^n \notin \mathcal{S}_n\}, \quad (53)$$

which yields (34) with (37).

On the other hand, from (39), the error probability satisfies

$$\varepsilon_n \le 1 - \frac{1}{M_n}\sum_{m \le M_n}(1-\lambda_n) = \lambda_n. \quad (54)$$

$\square$

*Proof of Theorem 2:* The converse part is omitted because it is proved by following the same lines as Theorem 1.

Let

$$R_0 \triangleq \sup_{\boldsymbol{X}:\bar{c}(\boldsymbol{Y})\le\Gamma} \underline{I}_\varepsilon(\boldsymbol{X};\boldsymbol{Y}) \quad (55)$$

and consider arbitrary $R < R_0$. For some $\gamma > 0$, there exists an $\boldsymbol{X}$ such that

$$R + \gamma < \underline{I}_\varepsilon(\boldsymbol{X};\boldsymbol{Y}), \quad (56)$$
$$\bar{c}(\boldsymbol{Y}) \le \Gamma, \quad (57)$$

where $\boldsymbol{Y} = \boldsymbol{Y}(\boldsymbol{X})$. From the definition of $\overline{c}(\boldsymbol{Y})$, we have

$$\lim_{n\to\infty} \Pr\left\{\frac{1}{n}c_n(Y^n) > \Gamma + \frac{1}{k}\right\} = 0 \quad (58)$$

for any natural number $k$. Here, we introduce an abbreviated notation:

$$\mathcal{Y}_{>\theta}^n \triangleq \left\{\boldsymbol{y} \in \mathcal{Y}^n \,\middle|\, \frac{1}{n}c_n(\boldsymbol{y}) > \theta\right\}. \quad (59)$$

Then, (58) can be written as

$$\Pr\left\{Y^n \in \mathcal{Y}_{>\Gamma+\frac{1}{k}}^n\right\} = \mathbb{E}\left[W^n\left(\mathcal{Y}_{>\Gamma+\frac{1}{k}}^n \,\middle|\, X^n\right)\right] \to 0, \quad (60)$$

which yields, from $W^n(\cdot|\cdot) \geq 0$,

$$\text{p-}\limsup_{n\to\infty} W^n\left(\mathcal{Y}_{>\Gamma+\frac{1}{k}}^n \,\middle|\, X^n\right) = 0. \quad (61)$$

Therefore, we have

$$\Pr\left\{W^n\left(\mathcal{Y}_{>\Gamma+\frac{1}{k}}^n \,\middle|\, X^n\right) > \frac{1}{k}\right\} < \frac{1}{k}, \quad (62)$$

for all $n$ large enough. With the diagonal line argument, we can show that there exists a vanishing series $\gamma_n \to 0$ satisfying

$$\Pr\left\{W^n\left(\mathcal{Y}_{>\Gamma+\gamma_n}^n \,\middle|\, X^n\right) > \gamma_n\right\} < \gamma_n. \quad (63)$$

We define

$$\mathcal{S}_n \triangleq \left\{\boldsymbol{x} \in \mathcal{X}^n \,\middle|\, W^n\left(\mathcal{Y}_{>\Gamma+\gamma_n}^n \,\middle|\, \boldsymbol{x}\right) \leq \gamma_n\right\} \quad (64)$$

and apply Lemma 2. Then, we have a code $(\varphi_n, \psi_n)$ satisfying

$$\frac{1}{n}\log M_n \geq R, \quad (65)$$

$$\varepsilon_n \leq \Pr\left\{\frac{1}{n}\log\frac{W^n(Y^n|X^n)}{P_{Y^n}(Y^n)} < R + \gamma\right\}$$
$$+ \Pr\{X^n \notin \mathcal{S}_n\} + e^{-n\gamma}. \quad (66)$$

From (56), the first term on the right hand side of (66) is bounded as

$$\limsup_{n\to\infty} \Pr\left\{\frac{1}{n}\log\frac{W^n(Y^n|X^n)}{P_{Y^n}(Y^n)} < R + \gamma\right\} \leq \varepsilon. \quad (67)$$

From (63) and (64), the second term vanishes. Hence we have

$$\limsup_{n\to\infty} \varepsilon_n \leq \varepsilon. \quad (68)$$

Furthermore, to check the requirement to the cost, we evaluate

$$\Pr\left\{\frac{1}{n}c_n(Y^n(\varphi_n(U_{M_n}))) > \Gamma + \gamma_n\right\} \quad (69)$$

$$= \frac{1}{M_n}\sum_{m\in\mathcal{M}_n} \Pr\left\{\frac{1}{n}c_n(Y^n(\varphi_n(m))) > \Gamma + \gamma_n\right\} \quad (70)$$

$$= \frac{1}{M_n}\sum_{m\in\mathcal{M}_n} W^n\left(\mathcal{Y}_{>\Gamma+\gamma_n}^n \,\middle|\, \varphi_n(m)\right) \quad (71)$$

$$\leq \frac{1}{M_n}\sum_{m\in\mathcal{M}_n} \gamma_n = \gamma_n \to 0, \quad (72)$$

where the inequality comes from $\varphi_n(m) \in \mathcal{S}_n$. Then, we have

$$\Gamma \geq \text{p-}\limsup_{n\to\infty} \left(\frac{1}{n}c_n(Y^n(\varphi_n(U_{M_n}))) - \gamma_n\right) \quad (73)$$

$$= \text{p-}\limsup_{n\to\infty} \frac{1}{n}c_n(Y^n(\varphi_n(U_{M_n}))). \quad (74)$$

From (65), (68), and (74), we can conclude that rate $R$ is $(\varepsilon, \Gamma)$-achievable and that $C(\varepsilon, \Gamma) \geq R_0$. $\qquad\square$

## IV. ALLOWING SMALL COST OVERRUN

Let us weaken the requirement for the cost constraint in the following way.

*Definition 8:* For $\boldsymbol{X}$ and $\delta \geq 0$, we define

$$\overline{c}_\delta(\boldsymbol{X}) \triangleq \inf\left\{\theta \,\middle|\, \text{p-}\limsup_{n\to\infty} W^n\left(\mathcal{Y}_{>\theta}^n|X^n\right) \leq \delta\right\}, \quad (75)$$

where $\mathcal{Y}_{>\theta}^n$ is defined by (59).

Note that $\overline{c}_\delta(\boldsymbol{X})$ is a generalization of $\overline{c}(\boldsymbol{Y}(\boldsymbol{X}))$. If $\delta = 0$, then, in view of (75),

$$\text{p-}\limsup_{n\to\infty} W^n\left(\mathcal{Y}_{>\theta}^n|X^n\right) \leq 0 \quad (76)$$

means

$$\lim_{n\to\infty} \Pr\left\{\frac{1}{n}c_n(Y^n(X^n)) > \theta\right\} = 0. \quad (77)$$

Then, we have

$$\overline{c}_0(\boldsymbol{X}) = \overline{c}(\boldsymbol{Y}(\boldsymbol{X})) = \text{p-}\limsup_{n\to\infty} \frac{1}{n}c_n(Y^n(X^n)). \quad (78)$$

*Definition 9:* Rate $R$ is said to be $(\varepsilon, \delta, \Gamma)$-achievable if and only if there exists a sequence of codes $\{(\varphi_n, \psi_n)\}_{n=1}^\infty$ satisfying

$$\lim_{n\to\infty} \varepsilon_n \leq \varepsilon, \quad (79)$$

$$\liminf_{n\to\infty} \frac{1}{n}\log M_n \geq R, \quad (80)$$

$$\overline{c}_\delta(\{\varphi_n(U_{M_n})\}_{n=1}^\infty) \leq \Gamma. \quad (81)$$

*Definition 10:* For $\varepsilon$, $\delta$, $\Gamma$, we define

$$C(\varepsilon, \delta, \Gamma) \triangleq \sup\{R|R \text{ is } (\varepsilon, \delta, \Gamma)\text{-achievable}\} \quad (82)$$

and call it $(\varepsilon, \delta, \Gamma)$-channel capacity.

We have the following theorem.

*Theorem 3:* $(\varepsilon, \delta, \Gamma)$-channel capacity is given by

$$C(\varepsilon, \delta, \Gamma) = \sup_{\boldsymbol{X}:\overline{c}_\delta(\boldsymbol{X})\leq\Gamma} \underline{I}_\varepsilon(\boldsymbol{X}; \boldsymbol{Y}). \quad (83)$$

*Proof:* The converse part is omitted because it is proved by following the same lines as Theorem 1.

Let

$$R_0 \triangleq \sup_{\boldsymbol{X}:\overline{c}_\delta(\boldsymbol{X})\leq\Gamma} \underline{I}_\varepsilon(\boldsymbol{X}; \boldsymbol{Y}) \quad (84)$$

and consider arbitrary $R < R_0$. For some $\gamma > 0$, there exists an $\boldsymbol{X}$ such that

$$R + \gamma < \underline{I}_\varepsilon(\boldsymbol{X}; \boldsymbol{Y}), \quad (85)$$

$$\overline{c}_\delta(\boldsymbol{X}) \leq \Gamma. \quad (86)$$

From the definition of $\overline{c}_\delta(\boldsymbol{X})$, we have

$$\text{p-}\limsup_{n\to\infty} W^n\left(\mathcal{Y}^n_{>\Gamma+\frac{1}{k}}\,\Big|\,X^n\right)\le\delta \qquad (87)$$

for any natural number $k$. This implies that

$$\lim_{n\to\infty}\Pr\left\{W^n\left(\mathcal{Y}^n_{>\Gamma+\frac{1}{k}}\,\Big|\,X^n\right)>\delta+\frac{1}{k}\right\}=0, \qquad (88)$$

and, for all $n$ large enough,

$$\Pr\left\{W^n\left(\mathcal{Y}^n_{>\Gamma+\frac{1}{k}}\,\Big|\,X^n\right)>\delta+\frac{1}{k}\right\}<\frac{1}{k}. \qquad (89)$$

With the diagonal line argument, we can show that there exists a vanishing series $\gamma_n\to 0$ safisfying

$$\Pr\left\{W^n\left(\mathcal{Y}^n_{>\Gamma+\gamma_n}\,\big|\,X^n\right)>\delta+\gamma_n\right\}<\gamma_n. \qquad (90)$$

We define

$$\mathcal{S}_n\triangleq\left\{\boldsymbol{x}\in\mathcal{X}^n\,\big|\,W^n\left(\mathcal{Y}^n_{>\Gamma+\gamma_n}\,\big|\,\boldsymbol{x}\right)\le\delta+\gamma_n\right\} \qquad (91)$$

and apply Lemma 2. Then, we have a code $(\varphi_n,\psi_n)$ satisfying

$$\frac{1}{n}\log M_n\ge R, \qquad (92)$$

$$\varepsilon_n\le\Pr\left\{\frac{1}{n}\log\frac{W^n(Y^n|X^n)}{P_{Y^n}(Y^n)}<R+\gamma\right\}$$
$$+\Pr\left\{X^n\notin\mathcal{S}_n\right\}+e^{-n\gamma}. \qquad (93)$$

From (85), the first term on the right hand side of (93) is bounded as

$$\limsup_{n\to\infty}\Pr\left\{\frac{1}{n}\log\frac{W^n(Y^n|X^n)}{P_{Y^n}(Y^n)}<R+\gamma\right\}\le\varepsilon. \qquad (94)$$

From (90) and (91), the second term vanishes. Hence we have

$$\limsup_{n\to\infty}\varepsilon_n\le\varepsilon. \qquad (95)$$

Furthermore, we check the requirement to the cost. Consider any $\gamma'>0$. Since $\gamma'>\gamma_n$ for all $n$ large enough, we have

$$\mathcal{Y}^n_{>\Gamma+\gamma'}\subset\mathcal{Y}^n_{>\Gamma+\gamma_n}, \qquad (96)$$

which yields

$$W^n(\mathcal{Y}^n_{>\Gamma+\gamma'}|\boldsymbol{x})\le W^n(\mathcal{Y}^n_{>\Gamma+\gamma_n}|\boldsymbol{x}) \qquad (97)$$

for all $\boldsymbol{x}\in\mathcal{X}^n$. Since codeword $\tilde{X}^n\triangleq\varphi_n(U_{M_n})$ is in $\mathcal{S}_n$, we have

$$\text{p-}\limsup_{n\to\infty}W^n(\mathcal{Y}^n_{>\Gamma+\gamma'}|\tilde{X}^n) \qquad (98)$$

$$\le\text{p-}\limsup_{n\to\infty}W^n(\mathcal{Y}^n_{>\Gamma+\gamma_n}|\tilde{X}^n) \qquad (99)$$

$$\le\limsup_{n\to\infty}(\delta+\gamma_n)=\delta. \qquad (100)$$

and therefore

$$\overline{c}_\delta(\tilde{\boldsymbol{X}})\le\Gamma+\gamma'. \qquad (101)$$

Since $\gamma'>0$ is arbitrary, we can derive

$$\overline{c}_\delta(\tilde{\boldsymbol{X}})\le\Gamma. \qquad (102)$$

From (92), (95), and (102), we can conclude that rate $R$ is $(\varepsilon,\delta,\Gamma)$-achievable and that $C(\varepsilon,\delta,\Gamma)\ge R_0$. $\qquad\square$

Here, we give a brief consideration to our relaxed constraint on the cost.

*Proposition 1:* For any $\boldsymbol{X}$ and $\boldsymbol{Y}=\boldsymbol{Y}(\boldsymbol{X})$,

$$\overline{c}_\delta(\boldsymbol{X})\ge\inf\left\{\theta\,\Big|\,\limsup_{n\to\infty}\Pr\left\{Y^n\in\mathcal{Y}^n_{>\theta}\right\}\le\delta\right\}. \qquad (103)$$

*Proof:* For fixed $\theta$, let

$$Z_n\triangleq W^n\left(\mathcal{Y}^n_{>\theta}|X^n\right). \qquad (104)$$

Since we can show that $\{Z_n\}_{n=1}^\infty$ satisfies the uniform integrability, it holds that [4, Lemma 5.3.2]

$$\text{p-}\limsup_{n\to\infty}Z_n\ge\limsup_{n\to\infty}\mathbb{E}[Z_n]. \qquad (105)$$

Then, for $\delta$,

$$\text{p-}\limsup_{n\to\infty}Z_n=\text{p-}\limsup_{n\to\infty}W^n\left(\mathcal{Y}^n_{>\theta}|X^n\right)\le\delta \qquad (106)$$

implies that

$$\limsup_{n\to\infty}\mathbb{E}[Z_n]=\limsup_{n\to\infty}\Pr\left\{Y^n\in\mathcal{Y}^n_{>\theta}\right\}\le\delta. \qquad (107)$$

From the definition of $\overline{c}_\delta(\boldsymbol{X})$, this means (103). $\qquad\square$

## V. General Cost

We have defined the cost function on the output alphabet of the channel. Generally, we can define the cost as a random variable correlated with the input of the channel. It should be noted that, in our situation, the cost does not have to take a real number. The set where the cost takes its value from can be absolutely arbitrary. It is enough to define a subset indicating a desired region. In other words, we can generalize the setting as follows.

Let $\Theta$ be a set which has no structure. As a cost, consider a random variable on $\Theta$ correlated with channel input $X^n$. Given a channel input $\boldsymbol{x}\in\mathcal{X}^n$, the probability of the cost is measured by $V_n(\cdot|\boldsymbol{x})$. For a desired region $\Gamma\subset\Theta$, defining

$$\overline{c}_\Gamma(\boldsymbol{X})\triangleq\text{p-}\limsup_{n\to\infty}V_n(\Gamma|X^n), \qquad (108)$$

we substitute (81) with

$$\overline{c}_\Gamma(\boldsymbol{X})\ge 1-\delta. \qquad (109)$$

Then, we can derive the following theorem with a proof similar to Theorem 3.

*Theorem 4:* $(\varepsilon,\delta,\Gamma)$-channel capacity is given by

$$C(\varepsilon,\delta,\Gamma)=\sup_{\boldsymbol{X}:\overline{c}_\Gamma(\boldsymbol{X})\ge 1-\delta}\underline{I}_\varepsilon(\boldsymbol{X};\boldsymbol{Y}). \qquad (110)$$

## VI. Conclusion

In this paper, we considered a cost defined on the output alphabet of the channel and derived channel coding theorems with cost constraint. In the fundamental setting, the cost constraint was represented by means of $\text{p-}\limsup$ because the cost was essentially a random variable. Moreover, we weakened the requirements for the error probability and for the cost constraint and extended the coding theorem. Finally, we introduced the notion of the general cost.

## ACKNOWLEDGMENT

This work was supported by JSPS KAKENHI Grant Number JP18K04131.

# Prefix Synchronized Codes and
# Random Access for DNA-Based Data Storage

Han Mao Kiah

School of Physical and Mathematical Sciences, Nanyang Technological University, Singapore

Email: hmkiah@ntu.edu.sg

## ABSTRACT

Prefix synchronized (PS) code, proposed by Gilbert [1], is a set of codewords that share the same prefix $p$ and whose coded data sequences avoid the prefix $p$. Even though sizes of PS-codes were studied by various authors, Morita et al. [2] provided an efficient method to encode binary messages into a maximal PS-code.

Recent advances in synthesis and sequencing technologies have made DNA macromolecules an attractive medium for digital information storage. Besides being biochemically robust, DNA strands offer ultrahigh storage densities of $10^{15} - 10^{20}$ bytes per gram of DNA, as demonstrated in recent experiments (see [3, Table 1]). However, this non-traditional media presents new challenges to the coding community (see [4] for a survey). One of which is *random access*.

To equip DNA-based data storage with random-access capabilities, Yazdi et al. [3], [5] developed an architecture that allows selective access to encoded DNA strands through the process of *hybridization*. Their technique involves prepending information-carrying DNA strands with specially chosen address sequences called *primers*. Yazdi et al. provided certain design considerations for these primers / prefixes [6] and also, verified the feasibility of their architecture in a series of experiments [3], [5]. As part of their proposal, Yazdi et al. extended the algorithm of Morita et al. [2] to encode data to avoid this set of primers / prefixes.

In this talk, we revisit the work of Morita et al. [2] and look at its connection to random access in DNA-based data storage. We also present related problems in the area such as design considerations and efficient constructions for these sets of primers [6], [7].

# Lossy Source Code Based on CoCoNuTS

Jun Muramatsu, NTT Communication Science Laboratories, NTT Corporation
Shigeki Miyake, NTT Network Innovation Laboratories, NTT Corporation

## Abstract

In the last workshop [16], we introduced the concept of CoCoNuTS (Codes based on Constrained Numbers Theoretically-achieving the Shannon limits), which provides building blocks for codes achieving the fundamental limits. We introduced a channel code based on this concept. In this workshop, we introduce a lossy source code based on the concept.

The idea of the construction is originated from the application of the Slepian-Wolf code to lossy source coding. In the construction [12], a lossless source code with decoder side information is used in the quantization function and a distributed lossless source code is used in the lossless compression function, where encoding and decoding were intractable. It should be noted that LDPC/LDGM codes were applied to lossy source coding [1], [4], [6], [7], [17] and the Wyner-Ziv problem [5], where it is assumed that source has a uniform distribution and the distortion function is the Hamming metric. In our series of papers, the achievability to the rate-distortion function is shown for arbitrary (non-uniform) source and additive distortion function.

At the same time, we had an alternative idea [8] that is similar to the nested-linear code [18], [19], where we use two functions (sparse matrices) to construct a code. We found that the essence of the quantization is *saturation property*, which comes from the hash property, of an ensemble of functions [13], where the results were extended to the Wyner-Ziv problem and the one-helps-one problem. However, encoding and decoding were still intractable. It should be noted that a similar idea is applied to the polar code in [2].

The first progress is using a variable-length code, which is used for the lossless compression function. The idea is introduced in [3]. This time, we obtained tractable algorithms using the sum-product algorithm for the lossless compression function. However, the quantization function was still intractable.

The second progress is using a stochastic quantizer called a *constrained-random-number generator* introduced in [9], [10]. The collision-resistance property is replaced with the *balanced-coloring property* introduced in [14]. Two approximation algorithms using the sum-product algorithm and the Markov Chain Monte Carlo method were developed. This time, we obtained tractable algorithms for the quantization function, which introduces a tractable algorithm for variable-length coding.

The last progress is using a constrained-random-number generator for the stochastic decoding in a fixed-length lossless compression. We have a tractable approximation algorithm instead of the maximum-likelihood decoder, which was intractable. This idea is introduced in [11], [15].